\documentclass[review,3p]{elsarticle}

\usepackage{hyperref}
\usepackage{wrapfig}

\journal{Neurocomputing}

\biboptions{sort&compress}
\begin{document}

\begin{frontmatter}

\title{Robust and Complex Approach of Pathological Speech Signal Analysis}

\author[but,six]{Jiri~Mekyska}
\author[iba]{Eva~Janousova}
\author[upm]{Pedro~Gomez-Vilda}
\author[but,six]{Zdenek~Smekal}
\author[fnusa,ceitec]{Irena~Rektorova\corref{cor}}
\ead{irena.rektorova@fnusa.cz}
\author[fnusa,ceitec]{Ilona~Eliasova}
\author[mu,ceitec]{Milena~Kostalova}
\author[fnusa,ceitec]{Martina~Mrackova}
\author[idetic]{Jesus~B.~Alonso-Hernandez}
\author[eupmt]{Marcos~Faundez-Zanuy}
\author[ehu]{Karmele~L\'{o}pez-de-Ipi\~{n}a}

\cortext[cor]{Corresponding author. Address: First Department of Neurology, St. Anne's University Hospital, Pekarska~53, 65691~Brno, Czech Republic. Telephone number: +420 549 497 825.}

\address[but]{Department of Telecommunications, Brno University of Technology, Technicka~10, 61600~Brno, Czech Republic}
\address[six]{SIX Research Centre, Technicka~10, 61600~Brno, Czech Republic}
\address[iba]{Institute of Biostatistics and Analyses, Masaryk University, Kamenice~3, 62500~Brno, Czech Republic}
\address[upm]{Facultad de Informatica, Universidad Politecnica de Madrid, Campus de Montegancedo, s/n, 28660~Boadilla del Monte, Madrid, Spain}
\address[fnusa]{First Department of Neurology, St. Anne's University Hospital, Pekarska~53, 65691~Brno, Czech Republic}
\address[ceitec]{Applied Neuroscience Research Group, Central European Institute of Technology, Masaryk University, Komenskeho~nam.~2, 60200 Brno, Czech Republic}
\address[mu]{Department of Neurology, Faculty Hospital and Masaryk University, Jihlavska~20, 63900~Brno, Czech Republic}
\address[idetic]{Institute for Technological Development and Innovation in Communications (IDeTIC), University of Las Palmas de Gran Canaria,
35001~Las Palmas de Gran Canaria, Spain}
\address[eupmt]{Escola Universitaria Politecnica de Mataro, Tecnocampus, Avda. Ernest Lluch~32, 08302~Mataro, Barcelona, Spain}
\address[ehu]{Department of Systems Engineering and Automation, University of the Basque Country UPV/EHU, Av de Tolosa~54, 20018 Donostia, Spain}

\begin{abstract}
This article presents a~study of the approaches in the state-of-the-art in the field of pathological speech signal analysis with a~special focus on parametrization techniques. It provides a~description of 92 speech features where some of them are already widely used in this field of science and some of them have not been tried yet (they come from different areas of speech signal processing like speech recognition or coding). As an original contribution, this work introduces 36 completely new pathological voice measures based on modulation spectra, inferior colliculus coefficients, bicepstrum, sample and approximate entropy and empirical mode decomposition. The significance of these features was tested on 3 (English, Spanish and Czech) pathological voice databases with respect to classification accuracy, sensitivity and specificity. To our best knowledge the introduced approach based on complex feature extraction and robust testing outperformed all works that have been published already in this field. The results (accuracy, sensitivity and specificity equal to $100.0\pm0.0\,\%$) are discussable in the case of Massachusetts Eye and Ear Infirmary (MEEI) database because of its limitation related to a~length of sustained vowels, however in the case of Pr{\'i}ncipe de Asturias (PdA) Hospital in Alcal{\'a} de Henares of Madrid database we made improvements in classification accuracy ($82.1\pm3.3\,\%$) and specificity ($83.8\pm5.1\,\%$) when considering a~single-classifier approach. Hopefully, large improvements may be achieved in the case of Czech Parkinsonian Speech Database (PARCZ), which are discussed in this work as well. All the features introduced in this work were identified by Mann-Whitney~U test as significant ($p < 0.05$) when processing at least one of the mentioned databases. The largest discriminative power from these proposed features has a~cepstral peak prominence extracted from the first intrinsic mode function ($p = 6.9443\cdot10^{-32}$) which means, that among all newly designed features those that quantify especially hoarseness or breathiness are good candidates for pathological speech identification. The article also mentions some ideas for the future work in the field of pathological speech signal analysis that can be valuable especially under the clinical point of view.
\end{abstract}


\end{frontmatter}

\section{Introduction}
Voice Pathology description and characterization has always demanded attention from the physiological and medical fields~\cite{Dejonckere2010}, as well as from the voice function point of view~\cite{Svec2008}. The term voice is described by~\cite{Titze1994} both in a~broad and a~narrow sense. In the broad sense voice may be taken as synonymous of speech, therefore terms as Voice over IP (VoIP) can be found in the literature and media with the meaning of speech data on internet. In the narrow sense voice refers to the vibration of the vocal folds. Speech sounds resulting from the interaction of this vibration with the Oro-Naso-Pharyngeal Tract (ONFT) are referred to, as voiced. Speech sounds produced by turbulent flow within the ONFT are termed voiceless. Phonation is the recommended term to refer to vocal fold vibration. A~speaker showing an anomalous vocal fold vibration pattern is referred as dysphonic, and aphonic if there is no vocal fold vibration at all. If no anomalies are present the speaker is referred as normophonic.

Dysphonic voice is a~perceptual and subjective term associated to Voice Pathology. Dysphonia is the perceptual quality of voice signaling that something wrong is happening in the phonation organs (mainly the larynx and its associated structures). Voice Pathologies and Dysphonic Voice are thus intrinsically related. The classification of Voice Pathologies or Disorders is described in~\cite{Titze1994} as tissue infection (e.\,g. laryngitis, bronchitis, croup,\dots), systemic changes (e.\,g. dehydratation, pharmacological and drug effects, hormonal changes,\dots), mechanical stress (e.\,g. vocal nodules, polyps, ulcers, granulomae, laryngocele, hemorrhage,\dots), surface irritation (e.\,g. laryngitis, leukoplakia, gastroesophageal reflux,\dots), tissue changes (e.\,g. laryngeal carcinoma, keratosis, papillomas, cysts,\dots), neurological and muscular changes (e.\,g. bilateral and unilateral vocal fold paralysis, Parkinson's Disease (PD), Amyotrophic Lateral Sclerosis (ALS), myotonic dystrophy, Huntington's Chorea, myasthenia gravis,\dots), and abnormal muscle patterns (e.\,g. conversion aphonia or dysphonia, spasmodic dysphonia, mutational dysphonia, ventricular phonation,\dots). In~\cite{Dworkin1997} a~description of vocal pathologies can be found. A~last important group of neurological diseases which leave a~correlate in voice and speech is that of cognitive origin, Alzheimer's Disease (AD) being the most relevant one for their impact in well-being and in aging specialized-attention demand. Some references on the influence of AD in speech and voice can be found in~\cite{Illes1989, Habash2012, Ipina2013, Horley2010}. Going a~step further, emotional alterations (either temporary or persistent) leave also correlates in the speech and phonation signature, and may be subjects of further study by acoustic analysis~\cite{Bucks2004, Gobl2003}

The relationship between acoustic correlates and voice pathology has been clinically established in the last decades, subjectively and quantitatively~\cite{Hirano1981, Baken2000}. Acoustic Voice Quality Analysis (AVQA) is a~wide term for a~set of different methodologies designed to quantify acoustic correlates giving a~definition of the quality of phonation or speech production. Therefore AVQA would be the procedural way to objectively quantify anomalies manifested in Dysphonic Voice. Modern signal processing technologies provide estimates of voice and speech correlates in time and frequency~\cite{Deller1993}, allowing to better visualize and quantify phonation patterns. Spectral techniques facilitate the study of pathologic phonation, establishing relations between harmonic-harmonic and harmonic-formant ratios, which were found as important correlates to organic voice pathology~\cite{Kuo1999}. Similarly, harmonic-noise ratios were found significant in characterizing certain types of dysphonic pathologies~\cite{Parsa2000, Murphy2007}. Time domain estimates, as jitter, shimmer and open-closed phase quotients are also used in describing dysphonic voice~\cite{Alku2003, Orr2003}. These descriptions gave rise to AVQA as a~specific field~\cite{Llorente2009}.

Under the point of view of AVQA the following objectives can be established in order of difficulty: dysphonic voice detection, dysphonic voice grading, and dysphonic voice classification according to etiology. Dysphonic voice detection would be the task of assigning normophonic (normative) or dysphonic (non-normative) labels to a~given phonation produced by a~specific speaker. The determination of the dysphonic grade is traditionally carried out by independent referees according to a~subjective criterion on a~given scale. One of the most popular is GRBAS (grade, roughness, breathiness, asthenia and strain)~\cite{Hirano1981}. The relative dependence of the assigned grade to the referee's subjective opinion, results in wide grading differences among referees. To overcome this problem requires the design of clinical assessment methodologies~\cite{Roy2013}. The task of dysphonic voice classification according to etiology is far more difficult, as a~given acoustic correlate may be attributed to different pathologies. If this problem is stated in terms of associating acoustic correlates to specific pathologies, the potential risk is that it will remain unsolved for long, because it is an ill-posed problem (a~many-to-one subjective mapping). A~preliminary step to be covered first is to define the implications of different pathologies in the vocal function, especially at the level of the larynx. Under the functional point of view, the following main behaviors may be observed in the abnormal operation of the vocal folds:  asymmetric vibration, contact defects and dystonia (hypo-, hyper-tension and tremor). Most organic larynx pathologies reproduce either one or another behavior, or all of them. Asymmetric vocal fold vibration is to be expected in pathologies as polyps, cysts, carcinomae, vocal fold paralysis, ulcers, cysts, papillomae, etc., and produces acoustic correlates as jitter, shimmer, poor harmonic-noise ratios, unbalance, sub- and inter-harmonics, etc. On its turn, contact defects are to be expected in pathologies as polyps, nodules, edemae, cysts, etc., where full closure of the glottal gap is not granted by vocal fold adduction and produce acoustic correlates as open and close phase perturbations, recovery phase attenuation, harmonic display reduction, etc. Other pathologies, especially those of neurological origin produce deviations in biomechanical parameters, as vocal fold tension (hypo- and hyper-tonic) and tremor, which can also be manifested as modulations in amplitude or frequency, and as changes in the vocal fold tension. As many organic pathologies induce a~hyper-tonic behavior, the main problem when dealing with correlates produced by neurological pathology is to differentiate their origin from that of organic origin. Contact defect pathologies can also show asymmetric vocal fold vibration, and this may also be the case in aging voice (presbyphonia). To establish differentiation criteria for using acoustic correlates when dealing with organic, neurologic or aging-induced perturbations, is a~major issue in AVQA~\cite{Vilda2013raey}.

Another problem regarding AVQA is the lack of good reference baselines to establish the methodologies for voice pathology classification from acoustic analysis. Rigorous databases, acquired to represent each of the different pathologies under well-defined sample population size and recording conditions are scant. Many times the problem comes from the simultaneous presence of different larynx pathologies in the same patient, either related or unrelated (e.g. it is common to find a~counter-lateral lesion as a~consequence of a~unilateral polyp). The problem then is how to decide if a~specific acoustic correlate is produced by one cause or another. Most of the times acoustic databases are produced by laryngological services as part of examination protocols~\cite{Saarbrucken2014, Repository2014}, but this is not always the case, as many laryngologists prefer to depend on visual exploration, neglecting the possibilities offered by AVQA for different reasons~\cite{Hakkesteegt2010}. Speech therapists rely more on acoustic exploration, but many times it is mainly restricted to measurements of vocal effort, long term frequency analysis, respiratory efficiency, or distortion measures. Therefore voice records from speakers ranked by etiology and severity index, using compatible standards (digitalization, channel, microphones) are scant. Most of the available databases are either incomplete, inconsistent (made up of recordings taken under different conditions) or deficient (many non-frequent pathologies are not well represented in sample size). Besides, there is a~lack of good normative databases, as most of the records produced by medical services contain information from dysphonic voice, but normative speakers are absent or poorly represented (this is the case of the most widely used database~\cite{MEEI}). This is a~severe limitation for systematic AVQA. Another problem is cross-lingual representation. As far as sustained vowels are concerned, this would not be a~problem, but it becomes a~major obstacle when segmental parameters are involved, as in the use of passages (either read or spontaneous).

After all these considerations, it may be said that the aim of AVQA is to design the best methodology to use Voice Correlates in Voice Pathology detection, grading and classification. Several steps are to be assumed and methodologically formulated for such. Voice Pathologies and Correlates have to be well connected by adequate physiological and biomechanical modeling of larynx dynamics to understand how acoustic correlates and organic dysfunctions could be associated~\cite{Titze2002}. This requirement is of vital importance in meeting the challenge of pathology classification from acoustic correlates. On its turn, inverse methods to obtain robust and significant estimates of voice correlates are of great relevance~\cite{Alku2011}. These are to be combined with other parameterization methods based in direct time and frequency domain estimates to produce rich feature sets~\cite{Llorente2006}. Finally, good Statistical Pattern Matching and Machine Learning methodologies have to be used to reduce redundancy and to make a selection by relevance within the feature set, to determine the best combinations for each specific task (detect, grade, classify), and procure specific results under quality criteria (sensitivity, specificity and accuracy)~\cite{Tsanas2010b}. These methods have to be contrasted on generally-accepted benchmark databases~\cite{Ghio2012}.

Having in mind all these conditions the present study is aimed to describe a~wide set of tests on different databases, using an exhaustive set of acoustic features, by means of well-known classifiers to estimate the performance of the methods and features regarding specificity, sensitivity and accuracy in dysphonic voice detection tasks. There are already many publications describing the methods of pathological speech detection. Usually the authors use just a~limited set of speech features and most of the publications present the detection accuracies tested only on one monolingual database~\cite{Lee2012, Vaziri2010, Markaki2010, Tsanas2010, Fredouille2009, Markaki2009, Little2007}. Probably the widest range of features describing different aspects of speech was tested in a~work of Tsanas et al.~\cite{Tsanas2010}. Regarding the databases there are just a~few publications considering 2 different data sets~\cite{Alpan2011, Vasilakis2009, Henriquez2009, Silva2009}.

To sum up the state-of-the-art approaches in the field of pathological voice analysis there is still a~lack of publications providing a~complex overview of features quantifying pathological speech and providing strong conclusions supported by a~robust testing. Therefore this work has 4 main goals:
\begin{enumerate}
	\item According to complex parameterization and consequent robust testing identify features that have the largest discriminative power in the field of pathological speech analysis.
	\item Design new features that can quantify hoarseness, breathiness and non-linearities in pathological speech signals.
	\item Prove that the proposed large set parameterization approach can provide better classification results (with respect to classification accuracy, sensitivity and specificity) than those published in the field of pathological speech analysis by the other researchers.
	\item Select a~database that has high potential for the future, especially in terms of speech features design, tuning and testing.
\end{enumerate}

The paper is organized as follows: the classically used features as well as the newly designed ones are discussed in section~\ref{sec:features}, section~\ref{sec:databases} describes the 3 databases, section~\ref{sec:experiments} describes the testing procedure, section~\ref{sec:results} is devoted to the experimental results and discussion. The conclusions are given in section~\ref{sec:conclusions}.

\section{Features}
\label{sec:features}

The aim of this work was to explore significance of the mostly used speech features when focusing on the ability of differentiation between healthy and pathological speech. The features that are usually known in different fields of speech signal processing (speech recognition, enhancement, denoising, identification etc.) and newly designed features originally introduced in this paper will be investigated as well.

Due to the limited size of this paper the features that were not originally introduced in this work will be mentioned without their deeper description by an algorithm. However each parameter will be accompanied by a~reference where the reader can find further information.

\subsection{Features Describing Phonation}
\label{sec:feat_phonation}

Probably the most popular features describing pathological voice are fundamental frequency $F_{0}$ and parameters describing its variability in time (jitter): PPQ5 (five-point Pitch Perturbation Quotient), RAP (Relative Average Perturbation), $\mbox{jitt}_{\mathrm{loc}}$ (average absolute difference between consecutive periods, divided by the average period), $\mbox{jitt}_{\mathrm{abs}}$ (average absolute difference between consecutive periods), $\mbox{jitt}_{\mathrm{ddp}}$ (average absolute difference between consecutive differences on neighbour glottal periods, divided by the average period)~\cite{Gelzinis2008, Vasilakis2009, Silva2009, Moers2011}. These features are good candidates especially for quantification of voice tremor~\cite{Skodda2010}.

A~disadvantage of previously mentioned measurements is that the values of the features are highly dependent on gender and variable acoustic environment. To overcome this disadvantage Little et al. proposed the PPE (Pitch Period Entropy)~\cite{Little2009}. During the calculation of PPE, logarithmic semitone scale, inverse filtering and entropy estimation are incorporated. Another method which is close to jitter is a~measure of standard deviation (std) of the time that vocal folds are apart ($\mbox{GQ}_{\mathrm{open}}$) and in collisions respectively ($\mbox{GQ}_{\mathrm{closed}}$)~\cite{Tsanas2010}.

To effectively describe hypophonia or intensity perturbations, short-time energy $E$~or pitch-level variations (shimmer) can be used~\cite{Tsanas2010, Gelzinis2008, Rektorova2007}. In this work 6 kinds of shimmer will be considered (they are calculated similarly to jitter but the intensity is used): APQ3 (three-point Amplitude Perturbation Quotient), APQ5, APQ11, $\mbox{shimm}_{\mathrm{loc}}$, $\mbox{shimm}_{\mathrm{ddp}}$, $\mbox{shimm}_{\mathrm{dB}}$ (average absolute base-10 logarithm of the difference between the amplitudes of consecutive periods)~\cite{Moers2011, Shao2009}.

Another feature describing speech intensity is TKEO (Teager-Kaiser Energy Operator)~\cite{Dimitriadis2009}. The advantage over simple $E$~is that it takes into account also signal frequency. It has been shown that speech contains dominant modulation frequencies in a~range 2--20\,Hz with maximum at approximately 4\,Hz~\cite{Falk2012}. The 4\,Hz modulation energy (ME) was selected as a~feature which is related to a~measure describing energy distribution in power spectra. Similarly MPSD (Median of Power Spectral Density), usually called median frequency, was selected~\cite{Gonzalez2010}. The last feature in this category is LSTER (Low Short-Time Energy Ratio)~\cite{Song2009}. This feature is usually used for differentiation between speech and music signals because speech exhibits higher variations in energy per 10--30\,ms frames. However this feature was selected for analysis of pathological speech as well. It is considered that patients will reach higher values of this feature in the case of maintained vowels due to the inability to sustain the same amount of airflow during the whole phonation.

\subsection{Features Describing Tongue Movement}

Frequencies of first three formants $F_{1}$, $F_{2}$, $F_{3}$ and their bandwidths $B_{1}$, $B_{2}$, $B_{3}$ are related to volumes of vocal tract cavities. Especially the volume of throat and oral cavity is modified by the tongue position. According to this fact it is possible to consider formants as a~measure of tongue movement~\cite{Weismer2001, Mekyska2011c}.

\subsection{Features Describing Speech Quality}

Signs of vocal fold dysfunctions are usually associated with breathiness or hoarseness~\cite{Vaziri2010}. From signal theory point of view these dysfunctions are characteristically decreasing voice quality, under a~simplified consideration, by additive noise. This implies that methods based on speech quality measurements can suitably quantify vocal folds impairment and describe its progress. Besides breathiness and hoarseness these methods can be also used for analysis of hypernasality caused by an improper work of soft palate~\cite{Orozco2011}.

Probably the simplest quality measure feature is ZCR (Zero-Crossing Rate) and its modification HZCRR (High Zero-Crossing Rate Ratio) which takes into account a~variation of ZCR in time~\cite{Song2009}. Another simple measure is FLUF (Fraction of Locally Unvoiced Frames) which can describe an impossibility of carrying out periodical glottal closure~\cite{Alonso2001}.

The next three measures are based on the variation of spectrum values between adjacent frames. Specifically these are SF (Spectral Flux)~\cite{Banchhor2012}, SDBM (Spectral Distance Based on Module) and SDBP (Spectral Distance Based on Phase)~\cite{Alonso2001}.

We have also used several features based on cepstral analysis. It was shown that cepstrum and its rahmonics correlate well with the perception of breathiness~\cite{Hillenbrand1996}. Moreover from the nature of cepstrum it can be said that it is a~kind of periodicity measure and therefore it should also predict roughness. The most famous feature based on the real cepstrum is CPP (Cepstral Peak Prominence) originally introduced by Hillenbrand et al.~\cite{Hillenbrand1996}. Besides this feature, PECM (Pitch Energy Cepstral Measure)~\cite{Alonso2001} and VR (Variation in Ratio between the second/first harmonic within the derived cepstral domain) were also used~\cite{Alonso2001}.

The other quality measure features are based on an estimation of the level of noise present. We will test the significance of HNR (Harmonic-to-Noise Ratio)~\cite{Orozco2011, Little2007, Alpan2011, Michaelis1997}, NHR (Noise-to-Harmonic Ratio)~\cite{Little2007, Deliyski1993}, NNE (Normalized Noise Energy)~\cite{Kasuya1986}, GNE (Glottal-to-Noise Excitation ratio)~\cite{Michaelis1997}, SPI (Soft Phonation Index)~\cite{Deliyski1993} and VTI (Voice Turbolence Index)~\cite{Deliyski1993}. The methods differ especially in the estimation of noise.

The last feature in this category is SSD (Segmental Signal-to-Dysperiodicity ratio). It was shown that this feature correlates strongly with the perceived degree of the speaker's hoarseness~\cite{Alpan2011}.

\subsection{Segmental Features}

Although some of the features mentioned in the other categories can be denoted as segmental as well (they are calculated from 10--30\,ms segments), here the segmental features are considered as matrices (not only vectors) calculated from the whole signal. Probably the most popular segmental features in the field of speech signal analysis are MFCC (Mel Frequency Cepstral Coefficients)~\cite{Gelzinis2008, Orozco2011, Markaki2009, Markaki2010}. The advantage of these coefficients is that they can indirectly detect slight misplacements of articulators~\cite{Tsanas2010}. In this work 20 MFCC coefficients are extracted. The coefficient number zero is replaced by an estimate of log-energy.

Although MFCC are generally used, there is a~lack of publications that compare these features to other segmental ones for the purpose of pathological speech analysis. Therefore we decided to include other segmental features. We extracted 20 mel frequency cepstral coefficients but in this case the bank of triangular filters was adjusted to the equal loudness curve~\cite{Makhoul1976}. We call these features MFCCE. The other two sets of features derived from MFCC are LFCC (Linear Frequency Cepstral Coefficients) and CMS (Cepstral Mean Subtraction coefficients). In the case of LFCC the bank of triangular filters is equidistantly spread in the frequency scale~\cite{Atassi2011}. CMS is a~kind of standardized z-score of MFCC (subtraction of mean and division by std over the time).

MSC (Modulation Spectra Coefficients) can provide information complementary to MFCC~\cite{Markaki2009, Atlas2003}. These features can capture a~class of source mechanism characteristics related to voice quality~\cite{Markaki2010}.

In the next step features based on linear prediction were extracted. LPC (Linear Predictive Coefficients)~\cite{Gelzinis2008}, PLP (Perceptual Linear Predictive coefficients)~\cite{Hermansky1990}, LPCC (Linear Predictive Cepstral Coefficients)~\cite{Mammone1996}, LPCT (Linear Predictive Cosine Transform coefficients)~\cite{Gelzinis2008} and ACW (Adaptive Component Weighted coefficients)~\cite{Mammone1996} were tested. The advantage of PLP over simple LPC or MFCC is that it also takes into account an adjustment to the equal loudness curve and intensity-loudness power law~\cite{Hermansky1990}. The advantage of LPCC and LPCT over ``classic'' LPC is that a~transformation is used into the cepstral domain and thus the values do not correlate much. The advantage of ACW is that these coefficients are less sensitive to channel distortion~\cite{Mammone1996}.

The last segmental features used in this work are parameters that analyze amplitude modulations in voice using a~biologically-inspired model of the inferior colliculus~\cite{Malyska2005}. These features are called ICC (Inferior Colliculus Coefficients).

Segmental features are sometimes extended by their $1^{\mathrm{st}}$ and $2^{\mathrm{nd}}$~order regression coefficients ($\Delta$ and $\Delta\Delta$ respectively). In this work used the $\Delta$ coefficients.

\subsection{Features Based on Bispectrum}

Alonso et al. proved that a~greater presence of quadratic coupling is observed in healthy voice when comparing it to the pathological one~\cite{Alonso2001}. It is probably due to a~fact that healthy voice is characterized by a~vocal tract which is more non-linear than in the case of pathological voices. This quadratic coupling can be appropriately described by bispectrum and features derived from this 2D signal.

There were proposed measures such as BII (Bicoherence Index Interference), HFEB (High Frequency Energy of one-dimensional Bicoherence), LFEB (low Frequency Energy of one-dimensional Bicoherence), BMII (Bispectrum Module Interference Index) and BPII (Bispectrum Phase Interference Index)~\cite{Alonso2001}.

\subsection{Features Based on Wavelet Decomposition}

The wavelet transform is widely used especially in the field of coding and speech denoising. However its application can be found in the field of pathological speech analysis too~\cite{Hosseini2008}. Detail coefficients after the decomposition can be used to estimate the present noise and consequently it is possible to calculate SNR (Signal-to-Noise ratio). In fact this is just another method of voice quality measurement.

According to some measurements we have empirically selected 7 wavelets: $10^{\mathrm{th}}$, $15^{\mathrm{th}}$, $20^{\mathrm{th}}$-order Daubechies wavelets; $10^{\mathrm{th}}$, $15^{\mathrm{th}}$, $20^{\mathrm{th}}$-order symlet wavelets and $5^{\mathrm{th}}$-order coiflet wavelet. We will mark the features derived from the detail coefficients of the wavelet transform as SNRW(\emph{wvl}), where \emph{wvl} corresponds to the specific wavelet (e.\,g. daub15).

\subsection{Features Based on Empirical Mode Decomposition}
\label{sec:feat_emd}

Recently, in speech processing new methods based on EMD (Empirical Mode Decomposition) have been used. Using EMD it is possible to decompose the arbitrary non-linear and time-varying signal into countable and usually a~small number of IMF (Intrinsic Mode Functions). These functions are modulated in amplitude and frequency and their sum gives the original signal.

Tsanas et al. proposed several measures of SNR and NSR based on the IMFs~\cite{Tsanas2010}. The time-varying high frequency components are present in the first few IMFs. Therefore these first few IMFs can be used to represent the noise in the signal and the rest of IMFs can be used for a~representation of the useful information. According to this consideration features like $\mbox{IMF-SNR}_\mathrm{TKEO}$ (based on Teager-Kaiser Energy Operator), $\mbox{IMF-SNR}_\mathrm{SEO}$ (based on Squared Energy Operator), $\mbox{IMF-SNR}_\mathrm{SE}$ (based on Shannon Entropy), $\mbox{IMF-NSR}_\mathrm{TKEO}$, $\mbox{IMF-NSR}_\mathrm{SEO}$ and $\mbox{IMF-NSR}_\mathrm{SE}$ have been introduced.

\subsection{Non-linear Dynamic Features}
\label{sec:feat_dynamic}

Tension on the vocal folds can significantly differ in the case of pathological speech. Voice becomes aperiodic, noisy-like, and it is very difficult to find any regularities in the signal. There is a~frequent presence of sub-harmonics and chaos which can lead to a~failure of conventional techniques of speech signal analysis. However it was shown that these kinds of signals can be sufficiently described by non-linear dynamical analysis~\cite{Little2007, Henriquez2009, Orozco2011, Vaziri2010}.

The first representative in this category is CD (Correlation Dimension) which statistically measures attractor geometry in the phase space. This measure is related to a~number of independent variables necessary for generating the attractor~\cite{Vaziri2010, Little2009, Shao2009}. Another dimension measure FD (Fractal Dimension) is based on a~number of basic building blocks that form a~pattern~\cite{Vaziri2010, Esteller2001}. We will also use complexity measures like ZL (Ziv-Lempel complexity) which quantify the regularity embedded in a~time series~\cite{Aboy2006}. Possible long-term dependencies in the speech signal will be described by HE (Hurst Exponent)~\cite{Orozco2011}.

Another set of measures is based on entropy. We will use SHE (Shannon Entropy), RE (second-order R{\'e}nyi Entropy)~\cite{Jayawardena2010}, CE (Correlation Entropy)~\cite{Jayawardena2010, Henriquez2009}, RBE1 (first-order R{\'e}nyi Block Entropy)~\cite{Henriquez2009}, RBE2 (second-order R{\'e}nyi Block Entropy)~\cite{Henriquez2009}, AE (Approximate Entropy)~\cite{Vaziri2010, Heris2009, Yentes2013}, SE (Sample Entropy)~\cite{Yentes2013} and RPDE (Recurrence Probability Density Entropy)~\cite{Little2007}. Generally the entropy is a~measure of uncertainty and it can be used to quantify the complexity of a~system. R{\'e}nyi entropies quantify the loss of information in time in a~dynamic system~\cite{Henriquez2009}, correlation entropy gives an indication of the predictability of the nonlinear time series~\cite{Jayawardena2010} and RPDE represents the uncertainty in the measurement of the pitch period~\cite{Little2007}. The only difference between AE and SE is that SE does not evaluate a~comparison of embedding vectors with themselves.

Another measure we have considered in this work is FMMI (First Minimum of Mutual Information function) which was found by Henriquez et al. as a~feature that better
discriminates among the different voice qualities of the multiquality database~\cite{Henriquez2009}. To include also a~measure of sensitivity to an initial condition, the LLE (Largest Lyapunov Exponent) was selected~\cite{Vaziri2010, Orozco2011}. We also used detrended fluctuation analysis (DFA) to characterize the self-similarity of the graph of a~signal from a~stochastic process~\cite{Little2007, Tsanas2010}. In this field NSE (Normalized Scaling Exponent) and FA (Fluctuation Amplitudes) were evaluated.

\subsection{High-level Features}

The features that are calculated for each speech segment separately form a~vector or a~matrix at the output of the parametrization process. This representation must be then transformed to a~scalar value to be able to carry out the next processing like statistical analysis, classification etc. This is usually done by an extraction of some kind of statistics. These statistics are called high-level features while parameters extracted directly from the speech signal are called local features. If the local feature is represented by a~matrix, then the high-level feature is calculated for each row separately (this monitors feature changes in time). We have extracted 60 high-level features:
\begin{itemize}
	\item max, min, position of max, position of min, relative pos. of max, relative pos. of min
	\item range, relative range, interquartile range, rel. interquartile range, interdecile range, rel. interdecile range, interpercentile range, rel. interpercentile range, studentized range
	\item mean, geometric mean, harmonic mean, mean excluding 10\,\%, 20\,\%, 30\,\%, 40\,\% and 50\,\%, of outliers, median, mode
	\item var, std, mean absolute deviation, median absolute dev., geometric standard dev., coefficient of variation, index of dispersion
	\item $3^{\mathrm{rd}}$, $4^{\mathrm{th}}$, $5^{\mathrm{th}}$ and $6^{\mathrm{th}}$ moment, kurtosis, skewness, Pearson's $1^{\mathrm{st}}$ and $2^{\mathrm{nd}}$ skewness coefficient
	\item $1^{\mathrm{st}}$, $5^{\mathrm{th}}$, $10^{\mathrm{th}}$, $20^{\mathrm{th}}$, $30^{\mathrm{th}}$, $40^{\mathrm{th}}$, $60^{\mathrm{th}}$, $70^{\mathrm{th}}$, $80^{\mathrm{th}}$, $90^{\mathrm{th}}$, $95^{\mathrm{th}}$ and $99^{\mathrm{th}}$ percentile, $1^{\mathrm{st}}$ and $3^{\mathrm{rd}}$ quartile
	\item slope, offset and error of linear regression
	\item modulation, Shannon entropy, second-order R{\'e}nyi entropy
\end{itemize}

\subsection{Newly Designed Features}

In sec.\,\ref{sec:feat_phonation} to sec.\,\ref{sec:feat_dynamic} we have presented 92 local features that are already used in the field of speech signal processing. However during the experiments these parameters will be extended by the other 36 features that are originally introduced in this work.

\subsubsection{Features Based on Modulation Spectra}

An initial step of modulation spectra calculation employs a~short-time Fourier transform (STFT) of the discrete input speech signal $s[n]$ with length $N$:
\begin{eqnarray}
S[k,m]&=&\sum\limits^{N-1}_{n = 0}s[n]w[n-mL]\mbox{e}^{-\mathrm{j}k\frac{2\pi}{N}n},\\
\nonumber
k&=&0,1,\dots,N-1,\\
\nonumber
m&=&0,1,\dots,M-1,
\end{eqnarray}
where $w[n]$ is a~window function (in our case a~Hamming window) with a~step of $L$ samples and $M$ number of speech segments. Consequently the power spectra $\left|S[k,m]\right|^{2}$ is extracted and filtered by a~bank of $P$ triangular filters equidistantly spaced out in the mel scale. This procedure forms a~matrix $X[p,m]$ with subbands $p = 1,2,\dots,P$. A~distribution of amplitudes of each subband envelope $X[p,m]$ of the voiced speech signal has a~strong exponential component which is suppressed by logarithmization and mean subtraction~\cite{Markaki2010}:
\begin{eqnarray}
\hat{X}[p,m] = \ln\left(X[p,m]\right) - \overline{\ln\left(X[p,m]\right)},
\end{eqnarray}
where $\overline{*}$ corresponds to the average operator over $m$. Next the frequency analysis of subband envelopes is performed using the discrete Fourier transform (DFT):
\begin{eqnarray}
\Psi[p,l]&=&\sum\limits^{M-1}_{m = 0}\hat{X}[p,m]\mbox{e}^{-\mathrm{j}l\frac{2\pi}{M}m},\\
\nonumber
l&=&0,1,\dots,M-1,
\end{eqnarray}
where $p$ and $l$ denote the acoustic and modulation frequency respectively. In the last step of modulation spectra extraction the second power of each subband is taken and normalized which partially suppresses the effect of training and testing conditions mismatch:
\begin{eqnarray}
\Psi_{\mathrm{n}}[p,l] = \frac{\Psi[p,l]}{\sum_{l}\Psi[p,l]}.
\end{eqnarray}

\begin{figure}[htbp]
	\centering
		\includegraphics{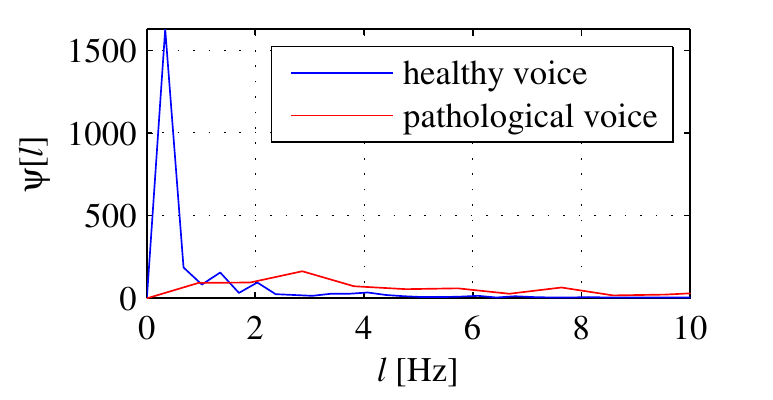}
	\caption{Function $\psi[l]$ calculated for healthy and pathological female vowels [a] obtained from the database PdA~\cite{Llorente2010, Londono2011}.}
	\label{fig:modulation_features}
\end{figure}

The features we are proposing are based on a function $\psi[l]$ which is extracted from the normalized modulation spectra according to:
\begin{eqnarray}
\psi[l] = \sum\limits^{P}_{p = 1}\Psi_{\mathrm{n}}[p,l].
\end{eqnarray}
Due to the instability of vocal fold vibrations pathological speech exhibits larger energy spread on higher modulation frequencies. This fact can be sufficiently expressed in function $\psi[l]$. Fig.\,\ref{fig:modulation_features} illustrates function $\psi[l]$ calculated for healthy and pathological female vowels [a] obtained from the database  Pr{\'i}ncipe de Asturias (PdA) Hospital in Alcal{\'a} de Henares of Madrid~\cite{Llorente2010, Londono2011}. As it can be seen the peak of function $\psi[l]$ in the case of healthy voice is higher, narrower and more shifted to lower modulation frequencies than in the case of the pathological one.

The proposed features derived from $\psi[l]$ are MSER (Modulation Spectra Energy Ratio), MFP (Modulation Frequency of Peak) and RPHM (Relative Peak Height of Modulation spectra). MSER is defined as:
\begin{eqnarray}
\mbox{MSER} = \frac{\sum\limits^{l_{\mathrm{5\,Hz}}}_{l = 0}\psi[l]}{\sum\limits^{M-1}_{l = l_{\mathrm{5\,Hz}}+1}\psi[l]},
\end{eqnarray}
where $l_{\mathrm{5\,Hz}}$ is a~sample corresponding to 5\,Hz modulation frequency (this limit was empirically found). MFP is defined as:
\begin{eqnarray}
\mbox{MFP} = \arg\max_{l}\left(\psi[l]\right),
\end{eqnarray}
where MFP is given in Hz. Finally RPHM can be calculated according to the following expression:
\begin{eqnarray}
\mbox{RPHM}&=&\frac{\psi[i]-r[i]}{\psi[i]},\\
\nonumber
i&=&\arg\max_{l}\left(\psi[l]\right),
\end{eqnarray}
where $r$ is a~linear regression line of $\psi[l]$ for $l = l_{\mathrm{5\,Hz}},\dots,M-1$ and $i$ is the index at which the function $\psi[l]$ reaches maximum value.

\subsubsection{Features Based on Inferior Colliculus Coefficients}

Using inferior colliculus coefficients (ICC) it is possible to extract the frequency content of the modulation envelopes applied to different bands of an auditory stimuli~\cite{Malyska2005}. Similarly to modulation spectra, the ICC extraction process employs at the beginning STFT and consequent calculation of power spectra $\left|S[k,m]\right|^{2}$. However in the next step instead of the bank of the triangular filters, the bank of $P$ mel-spaced gammatone filters is applied. We used filters defined by the the impulse response:
\begin{eqnarray}
g[n]&=&\left(\frac{n}{f_{\mathrm{s}}}\right)^{o-1}\cdot\cos\left(\frac{2\pi f_{\mathrm{c}}n}{f_{\mathrm{s}}}\right)\cdot\mbox{e}^{\frac{-2\pi bn}{f_{\mathrm{s}}}},\\
\nonumber
b&=&24.7\left(4.37^{-3}f_{\mathrm{c}}+1\right),
\end{eqnarray}
where $f_{\mathrm{c}}$ is the center frequency in Hz, $o$ the filter order (in our case 4) and $f_{\mathrm{s}}$ the sampling frequency in Hz. The DFT on the subband envelopes was applied next, this time without prior to normalization:
\begin{eqnarray}
T[p,l]&=&\sum\limits^{\infty}_{m = -\infty}X[p,m]\mbox{e}^{-\mathrm{j}l\frac{2\pi}{M}m}.
\end{eqnarray}
Finally the magnitude spectrum of each envelope $\left|T[p,l]\right|$ is filtered by a~bank of $Q = 13$ resonance filters with exponentially spaced frequencies from 12 to 107\,Hz. In this work we used resonance filters defined by the following transfer function:
\begin{eqnarray}
H(z) = \frac{0.1z^{2}-0.09}{z^{2}-1.8\cos\left(\frac{2\pi f_{\mathrm{c}}}{f_{\mathrm{s}}}\right)z+0.81}.
\end{eqnarray}
At the output of this process a~matrix $\Xi[p,q]$ ($q = 1,2,\dots,Q$) of ICC for the input signal $s[n]$ is extracted.

\begin{figure}[htbp]
	\centering
		\includegraphics{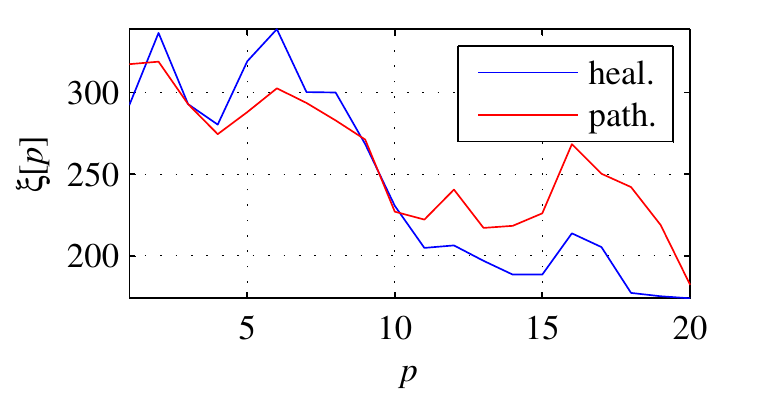}
	\caption{Function $\xi[p]$ calculated for the healthy and pathological female vowels [a] obtained from the database PdA~\cite{Llorente2010, Londono2011}.}
	\label{fig:ICC_features}
\end{figure}

The proposed features are based on a~function $\xi[p]$ which is extracted from $\Xi[p,q]$ according to:
\begin{eqnarray}
\xi[p] = \sum\limits^{Q}_{q = 1}\ln\left(\Xi[p,q]\right).
\end{eqnarray}
Contrary to $\psi[l]$ this function reflects the misplacement of articulators better than the instability of vocal fold vibrations. In Fig.\,\ref{fig:ICC_features} it is possible to see an example of $\xi[p]$ calculated for the pathological and healthy voice obtained from the PdA database. The features derived from $\xi[p]$ are ICER (Inferior Colliculus Energy Ratio) and RPHIC (Relative Peak Height of Inferior Colliculus). The ratio ICER is defined by the following expression:
\begin{eqnarray}
\mbox{ICER} = \frac{\sum\limits^{12}_{p = 1}\xi[p]}{\sum\limits^{20}_{p = 13}\xi[p]}.
\end{eqnarray}
Similarly to RPHM the RPHIC is extracted using the regression line but in this case $r[p]$ is calculated for $p = 13,\dots,20$:
\begin{eqnarray}
\mbox{RPHIC}&=&\frac{\xi[i]-r[i]}{\xi[i]},\\
\nonumber
i&=&\arg\max_{p}\left(\xi[p]\right).
\end{eqnarray}

\subsubsection{Features Based on Bicepstrum}

We will use a~definition of the real bicepstrum $c[n_{1},n_{2}]$ similar to the definition of the real cepstrum:
\begin{eqnarray}
c[n_{1},n_{2}]&=&\frac{1}{N^{2}}\sum\limits^{N-1}_{k_{1}, k_{2} = 0}\hat{B}[k_{1},k_{2}]\mbox{e}^{\mathrm{j}\frac{2\pi}{N}\left(k_{1}n_{1}+k_{2}n_{2}\right)},\\
\label{eq:logbispectrum}
\hat{B}[k_{1},k_{2}]&=&\ln\left(\left|B[k_{1},k_{2}]\right|\right),
\end{eqnarray}
where the bispectrum $B[k_{1},k_{2}]$ is calculated as the DFT of triple correlation or circular triple correlation $\gamma[n_{1},n_{2}]$~\cite{Kang1991}. In our work we used the second approach based on $\gamma[n_{1},n_{2}]$:
\begin{eqnarray}
B[k_{1},k_{2}]&=&\sum\limits^{N-1}_{n_{1}, n_{2} = 0}\gamma[n_{1},n_{2}]\mbox{e}^{-\mathrm{j}\frac{2\pi}{N}\left(k_{1}n_{1}+k_{2}n_{2}\right)},\\
\nonumber
\gamma[n_{1},n_{2}]&=&\frac{1}{N}\sum\limits^{N-1}_{n = 0}\delta[n]\delta[n+n_{1}]\delta[n+n_{2}],\\
\nonumber
\delta[i]&=&s[\left(n + i\right)~\mbox{mod}~N].
\end{eqnarray}
The first feature we are proposing in the present work is BCII (BiCepstral Index Interference):
\begin{eqnarray}
\mbox{BCII}&=&\frac{1}{N^{2}-1}\frac{1}{\left|\max\left(b[n_{1},n_{2}]\right)\right|}\cdot\\
\nonumber
&&\cdot\sum\limits^{N-1}_{n_{1} = 0}\sum\limits^{N-2}_{n_{2} = 0}\left|b[n_{1},n_{2}+1]-b[n_{1},n_{2}]\right|,\\
b[n_{1},n_{2}]&=&\frac{\left|\sum\limits^{M-1}_{m = 0}c_{m}[n_{1},n_{2}]\right|}{\sum\limits^{M-1}_{m = 0}\left|c_{m}[n_{1},n_{2}]\right|},
\end{eqnarray}
where $c_{m}[n_{1},n_{2}]$ is the bicepstrum calculated from the $m^{\mathrm{th}}$ speech segment. The two features introduced next are HFEBC (High Frequency Energy of one-dimensional BiCepstral index) and LFEBC (Low Frequency Energy of one-dimensional BiCepstral index):
\begin{eqnarray}
\mbox{LFEBC}&=&\frac{\sum\limits^{L}_{n = 0}\rho[n]}{\sum\limits^{N-1}_{n = 0}\rho[n]},\\
\mbox{HFEBC}&=&\frac{\sum\limits^{N-1}_{n = L+1}\rho[n]}{\sum\limits^{N-1}_{n = 0}\rho[n]},\\
\rho[n]&=&c[n,n],\\
\nonumber
L&=&\frac{f_{\mathrm{s}}}{f_{\mathrm{max}}},
\end{eqnarray}
where $\rho[n]$ (for~$n = 0,1,\dots,N-1$) is the one-dimensional bicepstral index and $f_{\mathrm{max}}$ the maximum expected fundamental frequency (in our case $f_{\mathrm{max}} = 350$\,Hz).

\begin{figure}[htbp]
	\centering
		\includegraphics{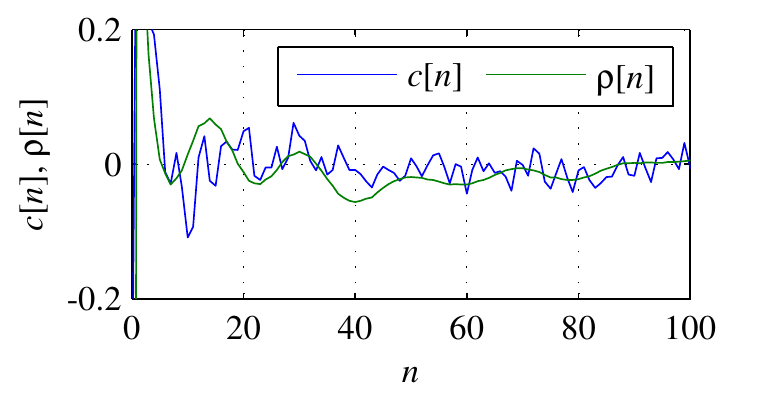}
	\caption{Comparison of the one-dimensional bicepstral index $\rho[n]$ and the real cepstrum $c[n]$ calculated for a~maintained vocal [a] uttered by a~healthy speaker. Only first 101 samples are displayed.}
	\label{fig:cepstrum_bicepstrum}
\end{figure}

It is supposed that pathological voice contains much more white noise (symmetrically distributed) than healthy voice. This is especially due to incorrect glottal closure. This kind of noise disappears in the real cepstrum estimated by $\rho[n]$ (see Fig.\,\ref{fig:cepstrum_bicepstrum}), therefore a~difference between the real cepstrum $c[n]$ (estimated using DFT) and the one-dimensional bicepstral index can estimate noise components of the analyzed signal. According to this idea BCMII (BiCepstrum Module Interference Index) and BCPII (BiCepstrum Phase Interference Index) are proposed:
\begin{eqnarray}
\mbox{BCMII}&=&\frac{1}{N^{2}-1}\frac{1}{\max\left(\eta_{\mathrm{m}}[m]\right)}\cdot\\
\nonumber
&&\cdot\sum\limits^{M-2}_{m = 0}\left|\eta_{\mathrm{m}}[m+1]-\eta_{\mathrm{m}}[m]\right|,\\
\mbox{BPMII}&=&\frac{1}{N^{2}-1}\frac{1}{\max\left(\eta_{\mathrm{p}}[m]\right)}\cdot\\
\nonumber
&&\cdot\sum\limits^{M-2}_{m = 0}\left|\eta_{\mathrm{p}}[m+1]-\eta_{\mathrm{p}}[m]\right|,\\
\eta_{\mathrm{m}}[m]&=&\sum\limits^{N-1}_{n = 0}\left(\left|c_{m}[n]\right|-\left|\rho_{m}[n]\right|\right)^{2},\\
\eta_{\mathrm{p}}[m]&=&\sum\limits^{N-1}_{n = 0}\left(\mbox{ang}\left(\tilde{c}_{m}[n]\right)-\mbox{ang}\left(\tilde{\rho}_{m}[n]\right)\right)^{2},
\end{eqnarray}
where $c_{m}[n]$ and $\rho_{m}[n]$ are the $m^{\mathrm{th}}$ frame real cepstrum and the one-dimensional bicepstral index respectively. $\tilde{c}_{m}[n]$ is the complex cepstrum and $\tilde{\rho}_{m}[n]$ is the one-dimensional bicepstral index where the absolute value in eq.\,(\ref{eq:logbispectrum}) was not taken.

The other features, based on $c[n]$ and $\rho[n]$ are LCBCER (Low Cepstra/BiCepstra Energy Ratio) and HCBCER (High Cepstra/BiCepstra Energy Ratio):
\begin{eqnarray}
\mbox{LCBCER}&=&\frac{\sum\limits^{L}_{n = 0}\left|c[n]\right|}{\sum\limits^{L}_{n = 0}\left|\rho[n]\right|},\\
\mbox{HCBCER}&=&\frac{\sum\limits^{N-1}_{n = L+1}\left|c[n]\right|}{\sum\limits^{N-1}_{n = L+1}\left|\rho[n]\right|}.
\end{eqnarray}
According to these equations we also propose the features LSBER (Low Spectra/Bispectra Energy Ratio) and HSBER (High Spectra/Bispectra Energy Ratio), however the ratios of spectrum $S[k]$ and one-dimensional bispectral index $\vartheta[k]$ for $L = \left\lfloor N/2\right\rfloor$ are calculated in this case.

The other two features BCMD (BiCepstral Module Distance) and BCPD (BiCepstral Phase Distance) are based on distance measures. The values of these features for the $m^{\mathrm{th}}$ speech segment are:
\begin{eqnarray}
\label{eq:BCMD}
\mbox{BCMD}&=&\sum\limits^{N-1}_{n_{1},n_{2} = 0}\left|\left|\tilde{c}_{m+1}[n_{1},n_{2}]\right|\right.-\\
\nonumber
&&-\left.\left|\tilde{c}_{m}[n_{1},n_{2}]\right|\right|,\\
\label{eq:BCPD}
\mbox{BCPD}&=&\sum\limits^{N-1}_{n_{1},n_{2} = 0}\left|\mbox{ang}\left(\tilde{c}_{m+1}[n_{1},n_{2}]\right)\right.-\\
\nonumber
&&-\left.\mbox{ang}\left(\tilde{c}_{m}[n_{1},n_{2}]\right)\right|,
\end{eqnarray}
where $\tilde{c}_{m}[n_{1},n_{2}]$ is the complex bicepstrum calculated without the absolute value in eq.\,(\ref{eq:logbispectrum}). Similarly to eq.\,(\ref{eq:BCMD}) and (\ref{eq:BCPD}) features BMD (Bispectral Module Distance) and BPD (Bispectral Phase Distance) are also extracted, where $B_{m}[k_{1},k_{2}]$ is used instead of $\tilde{c}_{m}[n_{1},n_{2}]$.

\subsubsection{Different Kernel Based Approximate and Sample Entropy}

Approximate entropy (AE) is a~measure of regularities inside the analyzed time series. The advantage of AE is that it can robustly estimate the system complexity using just a~limited number of samples (100--5000)~\cite{Chen2006}. To define AE we need to firstly reconstruct a~state space using Takens' embedding theorem:\,\cite{Takens1981}
\begin{eqnarray}
x[n]&=&\left[s[n],s[n+\tau],\dots,s[n+(m-1)\tau]\right],\\
\nonumber
n&=&0,1,\dots,N-1-(m-1)\tau,
\end{eqnarray}
where $x[n]$ is an embedding vector, $m$ is an embedding dimension and $\tau$ is a~time delay ($\tau = 1$ for AE). Next, the regularity quantity of a~particular pattern $C[i,m,r]$ is defined as:
\begin{eqnarray}
C[i,m,r] = \frac{1}{N-m}\sum\limits^{N-m}_{j = 0}\kappa\left(i,j,r\right).
\end{eqnarray}
Finally AE can be calculated according to expression:\,\cite{Vaziri2010}
\begin{eqnarray}
\mbox{AE}&=&\Phi[m,r]-\Phi[m+1,r],\\
\Phi[m,r]&=&\frac{1}{N-m}\sum\limits^{N-m}_{i = 0}\ln\left(C[i,m,r]\right).
\end{eqnarray}

A~disadvantage of AE is its dependence on the signal length due to the self comparison of points in the attractor. This fact can be avoided using the sample entropy (SE) which does not evaluate the comparison of embedding vectors among themselves:\,\cite{Orozco2013}
\begin{eqnarray}
\mbox{SE}&=&\Gamma[m,r]-\Gamma[m+1,r],\\
\Gamma[m,r]&=&\frac{1}{N-m}\sum\limits^{N-m}_{i = 0}\ln\left(C_{x}[i,m,r]\right),\\
C_{x}[i,m,r]&=&\frac{1}{N-m}\sum\limits^{N-m}_{j = 0, i \neq j}\kappa\left(i,j,r\right).
\end{eqnarray}
The usual values for embedding dimension are $m = 1,2$. In this work we used $m = 2$ which provides more detailed reconstruction. Finally, a~function must be defined $\kappa\left(i,j,r\right)$ we call kernel function. Originally the AE or SE are based on:
\begin{eqnarray}
\kappa\left(i,j,r\right)&=&\Theta\{r-d\left(x[i],x[j]\right)\},\\
d\left(x[i],x[j]\right)&=&\max_{k}|s[i+k]-s[j+k]|,\\
\nonumber
k&=&0,1,\dots,m-1,
\end{eqnarray}
where $\Theta$ is the Heaviside function and $r$ a~radius in our case calculated according to:
\begin{eqnarray}
r = 0.2\mbox{std}\left(s[n]\right).
\end{eqnarray}
We will denote these entropies as AE(Heaviside) and SE(Heaviside) respectively. Orozco-Arroyave et al. proposed AE(Gaussian) and SE(Gaussian) based on the Gaussian kernel:\,\cite{Orozco2013}
\begin{eqnarray}
\kappa\left(i,j,r\right) = \exp\left(-\frac{\|x[i]-x[j]\|^{2}}{10r^{2}}\right).
\end{eqnarray}
In this work we propose AE and SE based on the other 6 kernels: exponential kernel
\begin{eqnarray}
\kappa\left(i,j,r\right) = \exp\left(-\frac{\|x[i]-x[j]\|}{2r^{2}}\right);
\end{eqnarray}
Laplacian kernel,
\begin{eqnarray}
\kappa\left(i,j,r\right) = \exp\left(-\frac{\|x[i]-x[j]\|}{r}\right);
\end{eqnarray}
circular kernel,
\begin{eqnarray}
\kappa\left(i,j,r\right) = \frac{2}{\pi}\arccos\left(-\frac{\|x[i]-x[j]\|}{r}\right)-\\
\nonumber
-\frac{2}{\pi}\frac{\|x[i]-x[j]\|}{r}\sqrt{1-\left(\frac{\|x[i]-x[j]\|}{r}\right)^{2}},
\end{eqnarray}
for $\|x[i]-x[j]\| < r$, zero otherwise; spherical kernel,
\begin{eqnarray}
\kappa\left(i,j,r\right) = 1-\frac{3}{2}\frac{\|x[i]-x[j]\|}{r}+\\
\nonumber
+\frac{1}{2}{\left(\frac{\|x[i]-x[j]\|}{r}\right)}^{3},
\end{eqnarray}
for $\|x[i]-x[j]\| < r$, zero otherwise; Cauchy kernel,
\begin{eqnarray}
\kappa\left(i,j,r\right) = \frac{1}{1+\frac{\|x[i]-x[j]\|^{2}}{r}},
\end{eqnarray}
for $\|x[i]-x[j]\| < r$, zero otherwise and triangular kernel,
\begin{eqnarray}
\kappa\left(i,j,r\right) = 1-\frac{|x[i]-x[j]|}{r},
\end{eqnarray}
for $|x[i]-x[j]| < r$, zero otherwise.

\subsubsection{Features Based on Empirical Mode Decomposition}

As was mentioned in sec.\,\ref{sec:feat_emd} Tsanas et al. proposed several measures of SNR and NSR based on EMD:\,\cite{Tsanas2010}
\begin{eqnarray}
\mbox{IMF-SNR} = \frac{\sum\limits^{I}_{i = 4}\mu_{i}}{\sum\limits^{3}_{i = 1}\mu_{i}},\\
\mbox{IMF-NSR} = \frac{\sum\limits^{2}_{i = 1}\hat{\mu}_{i}}{\sum\limits^{I}_{i = 3}\hat{\mu}_{i}},
\end{eqnarray}
where $\mu_{i}$ is a~parameter, or mean sequence value, calculated from the original $i^{\mathrm{th}}$ IMF and $I$ is the total number of the IMFs. In the case of $\hat{\mu}$ the $i^{\mathrm{th}}$ IMF was logarithmized before the consequent parametrization. To extract the sequence from IMF, Tsanas et al. used SEO (Squared Energy Operator) and TKEO. As a~parameter they also used SHE.

We have extended this idea on these parameters: $\mbox{IMF-SNR}_\mathrm{RE}$ (based on second-order R{\'e}nyi Entropy), $\mbox{IMF-SNR}_\mathrm{ZCR}$ (based on Zero-Crossing Rate) and $\mbox{IMF-NSR}_\mathrm{RE}$. In the case of $\mbox{IMF-SNR}_\mathrm{RE}$ $\mu_{i}$ is defined as:
\begin{eqnarray}
\mu_{i} = -\log_{2}\left(\sum\limits^{J}_{j = 1}p^{2}\left(x^{i}_{j}\right)\right),
\end{eqnarray}
where $p^{2}\left(x^{i}_{j}\right)$ is the probability $P\left(\mbox{IMF}_{i} = x^{i}_{j}\right)$ and $\left\{x^{i}_{1}, x^{i}_{2}, \dots, x^{i}_{J}\right\}$ are the possible values of the $i^{\mathrm{th}}$ IMF. In the case of $\mbox{IMF-SNR}_\mathrm{ZCR}$ $\mu_{i}$ is calculated according to:
\begin{eqnarray}
\mu_{i}&=&\frac{1}{N}\sum\limits^{N-1}_{n = 1}\left|\mbox{sgn}\left(f_{i}[n]\right)-\mbox{sgn}\left(f_{i}[n-1]\right)\right|,
\end{eqnarray}
where $f_{i}[n]$ is the $i^{\mathrm{th}}$ IMF.

The time-varying high frequency components present in the $1^{\mathrm{st}}$ IMF represent the noise part of the speech signal. We propose a~new feature IMF-FD which is based on the fractal dimension calculated from the $1^{\mathrm{st}}$ IMF. This complexity measure appropriately quantifies the amount of noise present in the signal and indirectly describes hoarseness or breathiness. The feature is defined as:
\begin{eqnarray}
\mbox{IMF-FD}&=&\frac{\log_{10}N}{\log_{10}N + \log_{10}\left(\frac{N}{N+0.4N_{\mathrm{ch}}}\right)}\\
\nonumber
N_{\mathrm{ch}}&=&\sum\limits^{N-1}_{n = 1}\left|\mbox{sgn}\left(f_{1}[n]\right)-\mbox{sgn}\left(f_{1}[n-1]\right)\right|.
\end{eqnarray}

Another feature based on the first intrinsic mode function is IMF-CPP (Cepstral Peak Prominence extracted from the $1^{\mathrm{st}}$ IMF). It can be calculated as:
\begin{eqnarray}
\mbox{IMF-CPP}&=&\frac{c[i]-r[i]}{c[i]},\\
\nonumber
i&=&\arg\max_{n}\left(c[n]\right),\\
\nonumber
n&=&\frac{f_{\mathrm{s}}}{f_{\mathrm{max}}},\dots,N-1,
\end{eqnarray}
where $c[n]$ is real cepstrum calculated from $f_{1}[n]$ and $r[i]$ is a~linear regression line of $c[n]$ for $n = f_{\mathrm{s}}/f_{\mathrm{max}},\dots,N-1$. We consider $f_{\mathrm{max}} = 350\,\mbox{Hz}$.

The last feature proposed in this work is IMF-GNE (Glottal-to-Noise Excitation ratio based on the $1^{\mathrm{st}}$ IMF). The whole procedure of IMF-GNE extraction can be described in following steps:
\begin{enumerate}
	\item Segment the speech signal and extract the $1^{\mathrm{st}}$ IMF for each frame.
	\item Repeat step 3--6 for each segment.
	\item Do an inverse filtering of the $1^{\mathrm{st}}$ IMF.
	\item For each band of 1000\,Hz with 1000\,Hz step get the Hilbert envelope (the absolute value of analytical signal).
	\item Calculate cross correlation functions for all possible combinations of Hilbert envelopes and pick the maximum of each function.
	\item Pick the maximum from all the maxima in 5.
\end{enumerate}

\section{Databases}
\label{sec:databases}

To provide robust results 3 (English, Spanish and Czech) databases have been used during the testing procedure. Each database represents a~different language group (Germanic, Romanic and Slavic). This approach is advantageous from the cultural difference point of view. Speakers of different languages exhibit especially different prosodic characteristics. The aim of this work is to find features significant for the particular language, but we have focused on a~selection of features that are language-independent as well.

\subsection{MEEI Disordered Voice Database}

The Massachusetts Eye and Ear Infirmary (MEEI) database~\cite{MEEI} has been for many years used as a~benchmark in the field of pathological speech analysis. This commercially available database consists of 53 healthy and 657 pathological speakers with different pathologies (e.\,g. adductor spasmodic dysphonia, conversion dysphonia, erythema, hyperfunction, etc.). The data of each speaker contain 12\,s of a~standard text ``The Rainbow Passage''~\cite{Fairbanks1960} and a~sustained phonation of the vowel [a] pronounced as in the word ``father''. The recordings are sampled at $f_{\mathrm{s}} = 50\,\mbox{kHz}$ or $f_{\mathrm{s}} = 25\,\mbox{kHz}$. For our purpose only the vowels [a] are used.

Although MEEI consists of approximately 700 speakers in total, the age distributions between normal and pathological speakers are not matched. Therefore the amount of pathological speakers is usually limited to 173 according to criteria published by Parsa and Jamieson (we call this ``limited version of MEEI database'')~\cite{Parsa2000}. The statistical characteristics of MEEI database used in this work can be found in Table\,\ref{tab:MEEI}. 
\begin{table}[t!]
		\caption{Statistical characteristics of the MEEI database used in this work.}
		\label{tab:MEEI}
		\footnotesize
		\centering
		\begin{tabular}{l c c c c c c c c}
		\hline
		\hline
	  &\multicolumn{2}{c}{Number}&\multicolumn{2}{c}{Mean age}&\multicolumn{2}{c}{Age range}&\multicolumn{2}{c}{STD of age}\\
	  \cline{2-9}
		Speakers & Male & Female & Male & Female & Male & Female & Male & Female\\
		\hline
		Healthy & 21 & 32 & 38.81 & 34.16 & 26--59 & 22--52 & 8.49 & 7.87\\
		Pathological & 70 & 103 & 41.7 & 37.59 & 26--58 & 21--51 & 9.40 & 8.19\\
		\hline
		\hline
		\end{tabular}
\end{table}

Although this database is very popular and very often used, its size and data are considered insufficient. Table\,\ref{tab:MEEIresults} shows detection results (obtained on the MEEI database) of several works. The highest accuracy reached by Henriquez et al. is 99.69\,\%~\cite{Henriquez2009}. However using the same features and same experimental setup the authors achieved an accuracy of 82.47\,\% using the ``Multiquality'' database~\cite{Henriquez2009}. This fact shows that despite the popularity of the MEEI database there is a~need to introduce new, larger and more complex (when considering the speech tasks) benchmark databases in order to provide more reliable results and conclusions. A~discussion on the reliability of results mentioned in Table\,\ref{tab:MEEIresults} can be found in sec.\,\ref{sec:rule30}.
\begin{table}[h!]
		\caption{Summary of pathological speech detection results obtained on the MEEI database.}
		\label{tab:MEEIresults}
		\footnotesize
		\centering
		\begin{tabular}{l c}
		\hline
		\hline
		Reference & Accuracy [\%]\\
		\hline
		Henriquez et al. (2009)~\cite{Henriquez2009} & 99.69\\
		Diabazar et al. (2002)~\cite{Dibazar2002} & 99.44\\
		Parsa and Jamieson (2000)~\cite{Parsa2000} & 98.70\\
		Alpan et al. (2010)~\cite{Alpan2010} & 98.70\\
		Hariharan et al. (2010)~\cite{Hariharan2010} & 98.45\\
		Arias-Londono et al. (2011)~\cite{Londono2011b} & 98.23\\
		\hline
		\hline
		\end{tabular}
\end{table}

\subsection{PdA Database}
\label{sec:PdA}

The second database we have used is Pr{\'i}ncipe de Asturias (PdA) database~\cite{Llorente2010, Londono2011}. This database consists of 239 healthy and 200 pathological speakers with different organic pathologies (e.\,g. nodules, polyps, oedemas, carcinomas, etc.). Every speaker uttered a~sustained Spanish vowel [a]. The recordings are sampled at $f_{\mathrm{s}} = 25\,\mbox{kHz}$. The statistical characteristics of this database can be found in Table\,\ref{tab:PdA}. Table\,\ref{tab:PdAresults} shows detection results (obtained on the PdA database) of two available works. As can be seen, the accuracies are not so high as in the case of MEEI database. Moreover, the PdA consists of more speakers than the limited version of MEEI. All these facts highlight the high potential of the PdA for future use.
\begin{table}[t!]
		\caption{Statistical characteristics of the PdA database used in this work.}
		\label{tab:PdA}
		\footnotesize
		\centering
		\begin{tabular}{l c c c c c c c c}
		\hline
		\hline
	  &\multicolumn{2}{c}{Number}&\multicolumn{2}{c}{Mean age}&\multicolumn{2}{c}{Age range}&\multicolumn{2}{c}{STD of age}\\
	  \cline{2-9}
		Speakers & Male & Female & Male & Female & Male & Female & Male & Female\\
		\hline
		Healthy & 101 & 138 & 34.44 & 35.29 & 18--78 & 8--71 & 16.24 & 14.73\\
		Pathological & 74 & 126 & 48.05 & 36.71 & 11--76 & 9--72 & 13.89 & 13.14\\
		\hline
		\hline
		\end{tabular}
\end{table}
\begin{table}[ht!]
		\caption{Summary of pathological speech detection results obtained on the PdA database.}
		\label{tab:PdAresults}
		\footnotesize
		\centering
		\begin{tabular}{l c}
		\hline
		\hline
		Reference & Accuracy [\%]\\
		\hline
		Arias-Londono et al. (2011)~\cite{Londono2011}& 84.15\\
		Vasilakis and Stylianou (2009)~\cite{Vasilakis2009} & 77.68\\
		\hline
		\hline
		\end{tabular}
\end{table}

\subsection{PARCZ Database}
\label{sec:PARCZ}

The last database we have included in our test is the Czech Parkinsonian Speech Database (PARCZ) recorded at St. Anne's University Hospital in the Czech Republic. This database consists of 52 healthy speakers and 57 speakers with Parkinson's disease (PD) who suffer from hypokinetic dysarthria~\cite{Mekyska2011b}. This database contains 91 speech tasks (free speech, reading text, maintained vowels, diadochokinetic tasks, etc.) which are used for an analysis of speech dysfunctions that usually accompany PD. However for our purpose only the sustained Czech vowel [a] is used. The recordings are sampled at $f_{\mathrm{s}} = 48\,\mbox{kHz}$. The statistical characteristics of the PARCZ database can be found in Table\,\ref{tab:PARCZ}. As it can be seen, contrary to the MEEI or PdA, PARCZ is more focused on elder people.
\begin{table}[t!]
		\caption{Statistical characteristics of the PARCZ database used in this work.}
		\label{tab:PARCZ}
		\footnotesize
		\centering
		\begin{tabular}{l c c c c c c c c}
		\hline
		\hline
	  &\multicolumn{2}{c}{Number}&\multicolumn{2}{c}{Mean age}&\multicolumn{2}{c}{Age range}&\multicolumn{2}{c}{STD of age}\\
	  \cline{2-9}
		Speakers & Male & Female & Male & Female & Male & Female & Male & Female\\
		\hline
		Healthy & 26 & 26 & 65.65 & 62.15 & 49--83 & 45--87 & 9.02 & 9.50\\
		Pathological & 36 & 21 & 66.22 & 68.81 & 46--87 & 49--86 & 9.20 & 9.00\\
		\hline
		\hline
		\end{tabular}
\end{table}

\subsection{The Rule of 30}
\label{sec:rule30}

To decide whether the corpus size is sufficient for the robust conclusions Doddington et al. introduced ``the rule of 30'' which comes directly from the binomial distribution, assuming independent trials~\cite{Doddington2000}. The rule is ``To be 90\,\% confident that the true error rate is within $\pm30\,\%$ of the observed error rate, there must be at least 30 errors.''

If we apply this rule to the original MEEI database (710 speakers), then the observed error rates below 4.25\,\% cannot be considered as reliable. Moreover if the MEEI database is limited to 226 speakers, then threshold reaches 13.27\,\%. In other words, although the results in Table\,\ref{tab:MEEIresults} predict promising approaches for pathologic speech detection results, we must be critic about these values.

If we apply this rule to the PdA (436 speakers) or PARCZ (109 speakers) databases, we get the thresholds of 6.88\,\% and 27.52\,\% respectively. According to this values it can be said that the results obtained with the PdA are more reliable (contrary to PARCZ).

\section{Experiments}
\label{sec:experiments}

All the databases were resampled to $f_{\mathrm{s}} = 16\,\mbox{kHz}$ and in dependence on the next processing the data have been divided into 9 groups. Two approaches have been considered: gender-dependent and gender independent. Each database is randomly divided into 75\,\% and 25\,\% training and testing subsets respectively. The classifier is evaluated consequently. This procedure (data split, classifier tuning and evaluation) is repeated 100 times. The resulting accuracy ($ACC$), sensitivity ($SEN$) and specificity ($SPE$) are calculated according to: 
\begin{eqnarray}
ACC&=&\frac{TP+TN}{TP+TN+FP+FN}\cdot 100\,[\%],\\
SEN&=&\frac{TP}{TP+FN}\cdot 100\,[\%],\\
SPE&=&\frac{TN}{TN+FP}\cdot 100\,[\%],
\end{eqnarray}
where $TP$ (True Positive) and $FP$ (False Positive) represent the number of correctly identified pathological speakers and number of speakers diagnosed as pathological, but being healthy. Similarly, $TN$ (True Negative) and $FN$ (False Negative) represent the total number of correctly identified healthy speakers, and pathological speakers evaluated as healthy controls.

Before classification the training data are z-score normalized. The testing data are normalized by subtracting the training set mean and dividing by the training set standard deviation for each feature.

\subsection{Parameterization Programs}

Several toolboxes and programs have been used for the purpose of feature extraction. Features based on the detrended fluctuation analysis have been calculated using FastDFA~\cite{Little2006}. The glottal quotients ($\mbox{GQ}_{\mathrm{open}}$ and $\mbox{GQ}_{\mathrm{closed}}$) were extracted using the algorithm DYPSA implemented in VOICEBOX~\cite{Brookes2003}. To estimate the largest Lyapunov exponent (LLE) TSTOOL has been used~\cite{TSTOOL}. The software Praat has been used to estimate the fundamental frequency $F_{0}$, all kinds of jitter and shimmer, harmonic-to-noise ratio (HNR) and formant frequencies ($F_{1}, F_{2}, F_{3}$)~\cite{Praat}. The rest of features (108 in total), including the features introduced in this work, have been implemented in the Neurological Disorder Analysis Tool (NDAT) developed at the Brno University of Technology~\cite{Eliasova2013}.

\subsection{Feature Selection}

Considering all possible local and high-level feature combinations the parameterization process extracts approximately 28,000 features for each speaker. This feature space reduced for each scenario using the filtering feature selection approach based on the non-parametric Mann-Whitney U~test. The significance level was set to $\alpha = 0.05$. This feature selection method is relatively simple, but there are several serious studies indicating that there are scenarios where simple univariate methods perform similarly or even better than more complex methods. For instance Haury et al. show that a~simple Student's t-test provides better results than SVM-RFE (Support Vector Machine Recursive Feature Elimination), GFS (Greedy Forward Selection), LASSO (Least Absolute Shrinkage and Selection Operator) or elastic net~\cite{Haury2011}.

After the evaluation a~list of the ten most significant features is drawn up. Finally the density estimation plots (computed using kernel density estimation with Gaussian kernels) of the most significant features in all scenarios are given.

\subsection{Classification Methods}

In our test we have used 2 classifiers: SVM (Support Vector Machine with a~radial kernel) and RF (Random Forest). Regarding the SVM, the parameter kernel gamma $\gamma$ and penalty parameter $C$ were optimized using a~grid search for possible values.

\section{Results}
\label{sec:results}

A~summary of pathological speech detection results expressed by accuracy, sensitivity and specificity can be found in Table\,\ref{tab:results_all}. This table also provides some statistics related to the number of features selected in each scenario. In the case of MEEI database the accuracy, sensitivity and specificity were equal to $100.0\pm0.0\,\%$ in all scenarios (considering both genders together and separately) when using an RF classifier. A~discussion focused on the credibility of these results is given below. In comparison to RF the SVM classifier provided slightly worse results.

In the case of the PdA database the best results were found when classifying by RF too. The accuracy ($80.9\pm5.1\,\%$) and specificity ($85.4\pm6.7\,\%$) are larger for male speakers while the sensitivity ($77.2\pm7.7\,\%$) is larger for the female ones. When considering both genders together the accuracy ($82.1\pm3.3\,\%$) and sensitivity ($80.0\pm5.9\,\%$) reach the best values in the frame of all PdA scenarios, however the specificity is approximately 1.6\,\% lower than in the case of the male-only scenario. In comparison to the best accuracies published by Arias-Londono et al. our approach provides a~lower accuracy by 2.05\,\%, however it should be highlighted that the authors used classifier score fusion, which is not considered in this work~\cite{Londono2011}. When looking at their single-classifier solution, the authors reported these values: 81.70\,\% (accuracy), 80.50\,\% (sensitivity), 82.91\,\% (specificity). Comparing these values to our results, we can say that our approach provides better accuracy (by 0.40\,\%) and specificity (by 0.89\,\%), but worse sensitivity (by 0.50\,\%). Moreover, Arias-Londono et al. used in their work an older version of PdA database which has only 199 healthy speakers, while our version has 239. Therefore our results should be more trustable. In comparison to the work of Vasilakis et al. our approach outperformed the classification results provided by the authors~\cite{Vasilakis2009}.

Regarding the PARCZ database the results are much worse than in the case of MEEI or PdA. In scenario C1 (female speakers) the best accuracy was $67.1\pm8.3\,\%$ (classification by RF), sensitivity $35.3\pm30.3\,\%$ (SVM) and specificity $91.4\pm11.3\,\%$ (RF). The accuracy ($67.3\pm10.7\,\%$) and sensitivity ($50.4\pm20.2\,\%$), both obtained using the SVM, were slightly better in the case of male speakers. Vice versa, the specificity ($83.6\pm16.0\,\%$) was worse. Finally in scenario C3 (both genders) the best accuracy was $67.9\pm6.0\,\%$ (RF), sensitivity $39.3\pm14.9\,\%$ (SVM) and specificity $87.5\pm8.5\,\%$ (RF). It should be also mentioned that in comparison to MEEI or PdA the PARCZ database exhibits much larger standard deviations.

The poor classification results are caused probably by the fact that PD patients were in different progression stages of hypokinetic dysarthria, from first to more advanced ones (this fact most likely explains the large standard deviations that were obtained). It means that we were performing a~two-class classification over the data that can be split into approximately 4 classes (1 healthy and 3 levels of dysarthria: mild, moderate and severe). It can be an issue for future work to develop a~system that would not only identify the presence of pathological speech, but also estimate the level of voice pathology. The PARCZ database is a~good candidate for a~development of such a~system. The system can be interesting especially if we are able to detect the first stages of different disorders so that the doctors can start the treatment early and slow down the progress. One of the works that deals with this issue has been published by Henriquez et al.~\cite{Henriquez2009}

When looking into a~number of significant features selected by the Mann-Whitney U~test, it can be concluded that this number positively correlates with the classification accuracy. For example in the case of both genders the number of selected features is $15,521\pm231$ (MEEI), $11,540\pm419$ (PdA) and for PARCZ only $1,750\pm331$.

\begin{table}[ht!]
		\caption{Summary of pathological speech detection results represented as $\mbox{mean}\pm\mbox{std}$ [\%] (SVM\,--\,Support Vector Machine with a~radial kernel, RF\,--\,Random Forest, F\,--\,female, M\,--\,male, MF\,--\,all genders).}
		\label{tab:results_all}
		\scriptsize
		\centering
		\begin{tabular}{c c c c c c c c c c}
		\hline
		\hline
	  \multicolumn{3}{c}{Scenario}&No. of sel. features&\multicolumn{2}{c}{Accuracy}&\multicolumn{2}{c}{Sensitivity}&\multicolumn{2}{c}{Specificity}\\
	  \cline{5-10}
		ID & Dataset & Gender & ($\mbox{mean}\pm\mbox{std}$) & SVM & RF & SVM & RF & SVM & RF\\
		\hline
		M1 & MEEI & F & $13,996\pm288$ & $99.5\pm1.5$ & $\mathbf{100.0\pm0.0}$ & $99.3\pm2.0$ & $\mathbf{100.0\pm0.0}$ & $\mathbf{100.0\pm0.0}$ & $\mathbf{100.0\pm0.0}$\\
		M2 & MEEI & M & $13,561\pm398$ & $99.2\pm1.7$ & $\mathbf{100.0\pm0.0}$ & $99.1\pm2.1$ & $\mathbf{100.0\pm0.0}$ & $99.3\pm3.3$ & $\mathbf{100.0\pm0.0}$\\
		M3 & MEEI & MF & $15,521\pm231$ & $99.9\pm0.4$ & $\mathbf{100.0\pm0.0}$ & $99.8\pm0.5$ & $\mathbf{100.0\pm0.0}$ & $99.9\pm0.7$ & $\mathbf{100.0\pm0.0}$\\
		P1 & PdA & F & $9,726\pm447$ & $75.7\pm4.3$ & $\mathbf{78.5\pm4.9}$ & $72.8\pm6.5$ & $\mathbf{77.2\pm7.7}$ & $78.4\pm6.8$ & $\mathbf{79.6\pm7.3}$\\
		P2 & PdA & M & $9,721\pm458$ & $78.6\pm5.1$ & $\mathbf{80.9\pm5.1}$ & $71.0\pm10.1$ & $\mathbf{74.7\pm9.8}$ & $84.2\pm6.1$ & $\mathbf{85.4\pm6.7}$\\
		P3 & PdA & MF & $11,540\pm419$ & $77.7\pm3.2$ & $\mathbf{82.1\pm3.3}$ & $74.9\pm5.3$ & $\mathbf{80.0\pm5.9}$ & $80.1\pm5.0$ & $\mathbf{83.8\pm5.1}$\\
		C1 & PARCZ & F & $2,141\pm415$ & $65.9\pm11.9$ & $\mathbf{67.1\pm8.3}$ & $\mathbf{35.3\pm30.3}$ & $10.3\pm16.9$ & $79.0\pm15.3$ & $\mathbf{91.4\pm11.3}$\\
		C2 & PARCZ & M & $2,121\pm489$ & $\mathbf{67.3\pm10.7}$ & $66.5\pm10.3$ & $\mathbf{50.4\pm20.2}$ & $42.6\pm21.6$ & $79.4\pm14.5$ & $\mathbf{83.6\pm16.0}$\\
		C3 & PARCZ & MF & $1,750\pm331$ & $65.4\pm7.6$ & $\mathbf{67.9\pm6.0}$ & $\mathbf{39.3\pm14.9}$ & $31.0\pm14.6$ & $79.3\pm10.0$ & $\mathbf{87.5\pm8.5}$\\
		\hline
		\hline
		\end{tabular}
\end{table}

\begin{figure}
	\centering
		\includegraphics[width=1.00\textwidth]{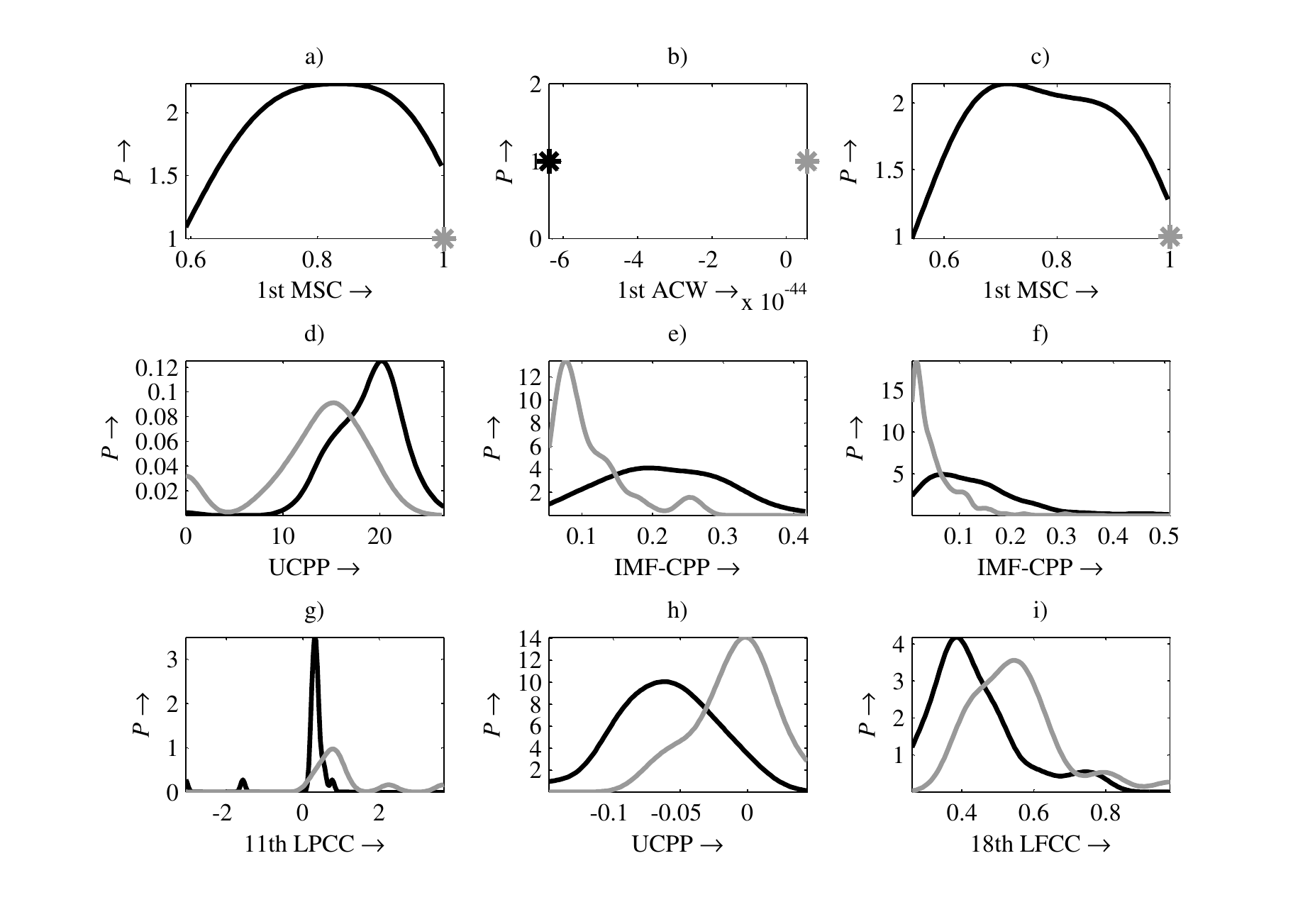}
	\caption{Density estimation plots (computed using kernel density estimation with Gaussian kernels) of the most significant features in scenarios M1\,--\,3, P1\,--\,3 and C1\,--\,3. Black colour represents healthy speech and grey the pathological one. a) Scenario M1 (MEEI, females): relative interpercentile range of 1st modulation spectra coefficients; b) Scenario M2 (MEEI, males): 3rd moment of 1st adaptive component weighted cepstral coefficients; c) Scenario M3 (MEEI, all): relative interpercentile range of 1st modulation spectra coefficients; d) Scenario P1 (PdA, females): mode of unsmooth cepstral peak prominence; e) Scenario P2 (PdA, males): median absolute deviation of cepstral peak prominence of first intrinsic mode function; f) Scenario P3 (PdA, all): error of linear regression of cepstral peak prominence of first intrinsic mode function; g) Scenario C1 (PARCZ, females): coefficient of variance of 11th linear predictive cepstral coefficients; h) Scenario C2 (PARCZ, males): slope of unsmooth cepstral peak prominence; i) Scenario C3 (PARCZ, all): interquartile range of 18th linear frequency cepstral coefficients.}
	\label{fig:distribution}
\end{figure}

The ten most significant features selected by Mann-Whitney~U test in all considered scenarios can be found in Tables\,\ref{tab:sign_M1}\,--\,\ref{tab:sign_C3}. The density estimation plots (computed using kernel density estimation with Gaussian kernels) of the most significant features in these scenarios can be seen in Fig.\,\ref{fig:distribution}. In all MEEI scenarios (M1\,--\,3) the 10 most significant features produce equivalent $p$ values and they are sorted alphabetically. Regarding scenarios M1 (females) and M3 (both genders) the most discriminative features are those derived from MSC (Modulation Spectra Coefficients) while in the case of M2 (males) features based on ACW (Adaptive Component Weighted coefficients) and FADFA (Fluctuation Amplitudes of Detrended Fluctuation Analysis) are mainly selected. Looking at Fig.\,\ref{fig:distribution} a)\,--\,c) it can be concluded that the relative interpercentile range of first modulation spectra coefficients and the third moment of the first adaptive component weighted cepstral coefficients show always a~single value for pathological speech. Moreover, in the case of male speakers the single value also represents the healthy one. In other words, regarding the MEEI database we can use a~very simple classifier (theoretically a~decision tree with only one node) to differentiate between healthy and pathological speech. This fact supports our classification accuracies equal to $100.0\pm0.0\,\%$. But the question is: Are these results trustable enough?

The MEEI database itself introduces many issues related to the credibility of the results. The main problem is that both the healthy and disordered speech were recorded in a~different way. First of all the healthy one was sampled at $f_{\mathrm{s}} = 50\,\mbox{kHz}$ and the disordered one at $f_{\mathrm{s}} = 25\,\mbox{kHz}$. In our experiment all databases were resampled to $f_{\mathrm{s}} = 16\,\mbox{kHz}$ but some small differences will remain among signals. But probably the most serious problem is that the duration of a~sustained vowel is 3\,s and 1\,s for healthy and disordered voice respectively. Therefore all high-level or local features that somehow reflect the signal length (entropy, range, std, duration, etc.) can provide very good discriminative results. This can also be a~case of the relative interpercentile range mentioned in Table\,\ref{tab:sign_M1} and \ref{tab:sign_M3}. Although it is a~relative measure it is still dependent on the length of the input vector. We made a~small experiment repetitively and randomly generating vectors with length $N = 10^{i}$ for $i = 1, 2, 3, 4$ and with a~normal distribution. After that we calculated the relative interpercentile range of these vectors and found out that the value of this measure is increasing with the decreasing vector length.

To avoid the problem of different vowel durations one can skip features dependent on this property but we would loose parameters that can be potentially very good candidates for pathological speech identification. The other approach is to take only a~one-second segment from the healthy speech but in our opinion this is not a~good solution. The first second segment could be selected but then we are losing the sustained part of the signal (used for estimating jitter) and phonation trail. If we take only the sustained part, then we will loose information about phonation onset and offset. It has been shown that an analysis of these parts is useful for example for the dysarthric speech description~\cite{Goberman2008, Eliasova2013}. Therefore there is no elegant solution that would not distort the results. Moreover, Malysla et al. mention that some speakers were recorded in different sites and over potentially different channels~\cite{Malyska2005}. In conclusion, although the MEEI database is a~very popular benchmark in the field of pathological speech analysis, the results obtained using this database should be taken very carefully. An introduction of a~new English disordered voice database is thus highly important for the future evaluation of new state-of-the-art speech signal processing techniques.

\begin{table}[ht!]
		\caption{10 most significant features selected by Mann-Whitney U~test in scenario M1: MEEI, females (MSC\,--\,Modulation Spectra Coefficients).}
		\label{tab:sign_M1}
		\footnotesize
		\centering
		\begin{tabular}{l l}
		\hline
		\hline
		Feature & $p$ value\\
		\hline
		relative interpercentile range of 1st MSC & $2.8610\cdot10^{-30}$\\
		relative interpercentile range of 10th MSC & $2.8610\cdot10^{-30}$\\
		relative interpercentile range of 11th MSC & $2.8610\cdot10^{-30}$\\
		relative interpercentile range of 12th MSC & $2.8610\cdot10^{-30}$\\
		relative interpercentile range of 13th MSC & $2.8610\cdot10^{-30}$\\
		relative interpercentile range of 14th MSC & $2.8610\cdot10^{-30}$\\
		relative interpercentile range of 15th MSC & $2.8610\cdot10^{-30}$\\
		relative interpercentile range of 16th MSC & $2.8610\cdot10^{-30}$\\
		relative interpercentile range of 17th MSC & $2.8610\cdot10^{-30}$\\
		relative interpercentile range of 18th MSC & $2.8610\cdot10^{-30}$\\
		\hline
		\hline
		\end{tabular}
\end{table}

\begin{table}[ht!]
		\caption{10 most significant features selected by Mann-Whitney U~test in scenario M2: MEEI, males (ACW\,--\,Adaptive Component Weighted coefficients, FADFA\,--\,Fluctuation Amplitudes of Detrended Fluctuation Analysis).}
		\label{tab:sign_M2}
		\footnotesize
		\centering
		\begin{tabular}{l l}
		\hline
		\hline
		Feature & $p$ value\\
		\hline
		3rd moment of 1st ACW & $2.5336\cdot10^{-21}$\\
		4th moment of 1st ACW & $2.5336\cdot10^{-21}$\\
		5th moment of 1st ACW & $2.5336\cdot10^{-21}$\\
		6th moment of 1st ACW & $2.5336\cdot10^{-21}$\\
		mean absolute deviation of 1st ACW & $2.5336\cdot10^{-21}$\\
		mean excluding 50\,\% outliers of 1st ACW & $2.5336\cdot10^{-21}$\\
		mean of 1st ACW & $2.5336\cdot10^{-21}$\\
		offset of linear regression of 1st ACW & $2.5336\cdot10^{-21}$\\
		position of max. of FADFA & $2.5336\cdot10^{-21}$\\
		relative position of min. of FADFA & $2.5336\cdot10^{-21}$\\
		\hline
		\hline
		\end{tabular}
\end{table}

\begin{table}[ht!]
		\caption{10 most significant features selected by Mann-Whitney U~test in scenario M3: MEEI, all (MSC\,--\,Modulation Spectra Coefficients).}
		\label{tab:sign_M3}
		\footnotesize
		\centering
		\begin{tabular}{l l}
		\hline
		\hline
		Feature & $p$ value\\
		\hline
		relative interpercentile range of 1st MSC & $1.0560\cdot10^{-49}$\\
		relative interpercentile range of 10th MSC & $1.0560\cdot10^{-49}$\\
		relative interpercentile range of 11th MSC & $1.0560\cdot10^{-49}$\\
		relative interpercentile range of 12th MSC & $1.0560\cdot10^{-49}$\\
		relative interpercentile range of 13th MSC & $1.0560\cdot10^{-49}$\\
		relative interpercentile range of 14th MSC & $1.0560\cdot10^{-49}$\\
		relative interpercentile range of 15th MSC & $1.0560\cdot10^{-49}$\\
		relative interpercentile range of 16th MSC & $1.0560\cdot10^{-49}$\\
		relative interpercentile range of 17th MSC & $1.0560\cdot10^{-49}$\\
		relative interpercentile range of 18th MSC & $1.0560\cdot10^{-49}$\\
		\hline
		\hline
		\end{tabular}
\end{table}

In scenario P1 (PdA, males) all the most significant features are based on UCPP (Unsmooth Cepstral Peak Prominence). On the other hand, in the case of scenarios P2 (females) and P3 (both genders) all the selected features are based on IMF-CPP (Cepstral Peak Prominence of first Intrinsic Mode Function) which is originally introduced in this work. However looking at Tables\,\ref{tab:sign_P1}\,--\,\ref{tab:sign_P3} it is evident that the features listed here correlate significantly (e.\,g. mean and median, std and var, etc.). We have not carried out an analysis of correlation in this work due to a~large number of features, but some statistics related to the most popular ones have been published, for example by Tsanas et al.~\cite{Tsanas2010b}

\begin{table}[ht!]
		\caption{10 most significant features selected by Mann-Whitney U~test in scenario P1: PdA, females (UCPP\,--\,Unsmooth Cepstral Peak Prominence).}
		\label{tab:sign_P1}
		\footnotesize
		\centering
		\begin{tabular}{l l}
		\hline
		\hline
		Feature & $p$ value\\
		\hline
		mode of UCPP & $1.6224\cdot10^{-19}$\\
		mean of UCPP & $2.9605\cdot10^{-19}$\\
		mean excluding 10\,\% outliers of UCPP & $3.6871\cdot10^{-19}$\\
		60th percentile of UCPP & $4.2979\cdot10^{-19}$\\
		mean excluding 20\,\% outliers of UCPP & $4.5230\cdot10^{-19}$\\
		mean excluding 40\,\% outliers of UCPP & $4.9362\cdot10^{-19}$\\
		median of UCPP & $4.9719\cdot10^{-19}$\\
		mean excluding 30\,\% outliers of UCPP & $5.0820\cdot10^{-19}$\\
		mean excluding 50\,\% outliers of UCPP & $5.0820\cdot10^{-19}$\\
		90th percentile of UCPP & $6.6980\cdot10^{-19}$\\
		\hline
		\hline
		\end{tabular}
\end{table}

\begin{table}[ht!]
		\caption{10 most significant features selected by Mann-Whitney U~test in scenario P2: PdA, males (IMF-CPP\,--\,Cepstral Peak Prominence of first IMF).}
		\label{tab:sign_P2}
		\footnotesize
		\centering
		\begin{tabular}{l l}
		\hline
		\hline
		Feature & $p$ value\\
		\hline
		median absolute deviation of IMF-CPP & $6.4974\cdot10^{-17}$\\
		interquartile range of IMF-CPP & $1.2600\cdot10^{-16}$\\
		60th percentile of IMF-CPP & $8.6433\cdot10^{-16}$\\
		3rd quartile of IMF-CPP & $1.1901\cdot10^{-15}$\\
		error of linear regression of IMF-CPP & $1.3454\cdot10^{-15}$\\
		median of IMF-CPP & $1.4129\cdot10^{-15}$\\
		mean absolute deviation of IMF-CPP & $2.0881\cdot10^{-15}$\\
		mean excluding 50\,\% outliers of IMF-CPP & $2.2461\cdot10^{-15}$\\
		70th percentile of IMF-CPP & $2.3014\cdot10^{-15}$\\
		mean excluding 40\,\% outliers of IMF-CPP & $3.1541\cdot10^{-15}$\\
		\hline
		\hline
		\end{tabular}
\end{table}

\begin{table}[ht!]
		\caption{10 most significant features selected by Mann-Whitney U~test in scenario P3: PdA, all (IMF-CPP\,--\,Cepstral Peak Prominence of first IMF).}
		\label{tab:sign_P3}
		\footnotesize
		\centering
		\begin{tabular}{l l}
		\hline
		\hline
		Feature & $p$ value\\
		\hline
		error of linear regression of IMF-CPP & $6.9443\cdot10^{-32}$\\
		median absolute deviation of IMF-CPP & $1.3082\cdot10^{-31}$\\
		mean absolute deviation of IMF-CPP & $1.8834\cdot10^{-31}$\\
		80th percentile of IMF-CPP & $4.5625\cdot10^{-31}$\\
		interquartile range of IMF-CPP & $5.3469\cdot10^{-31}$\\
		std. of IMF-CPP & $5.5387\cdot10^{-31}$\\
		var. of IMF-CPP & $5.5387\cdot10^{-31}$\\
		3rd quartile of IMF-CPP & $5.7880\cdot10^{-31}$\\
		70th percentile of IMF-CPP & $7.0861\cdot10^{-31}$\\
		interdecile range of IMF-CPP & $7.4045\cdot10^{-31}$\\
		\hline
		\hline
		\end{tabular}
\end{table}

In the case of scenario C1 (PARCZ, females) there were selected features based on LPCC (Linear Predictive Cepstral Coefficients), GNE (Glottal-to-Noise Excitation ratio), MPSD (Median of Power Spectral Density) and $\mbox{GQ}_{\mathrm{open}}$ (Qlottis Quotient\,--\,vocal folds are apart). Regarding scenario C2 (males), the 10 most significant features are based on UCPP, LPCT (Linear Predictive Cosine Transform coefficients), CMS (Cepstral Mean Subtraction coefficients), $\mbox{IMF-NSR}_\mathrm{RE}$ (Noise-to-Signal Ratio derived from IMF based on second-order R{\'e}nyi Entropy) and AE (Laplacian: Approximate Entropy based on Laplacian kernel). And finally, when considering both genders, there were selected features based on LFCC (Linear Frequency Cepstral Coefficients), MPSD, FADFA and ICC (Inferior Colliculus Coefficients). Generally $p$ values are much lower than in the case of MEEI or PdA databases. This is also closely related to the poor classification results mentioned in Table\,\ref{tab:results_all}.

\begin{table}[ht!]
		\caption{10 most significant features selected by Mann-Whitney U~test in scenario C1: PARCZ, females (LPCC\,--\,Linear Predictive Cepstral Coefficients, GNE\,--\,Glottal-to-Noise Excitation ratio, MPSD\,--\,Median of Power Spectral Density, $\mbox{GQ}_{\mathrm{open}}$\,--\,Qlottal Quotient (vocal folds are apart)).}
		\label{tab:sign_C1}
		\footnotesize
		\centering
		\begin{tabular}{l l}
		\hline
		\hline
		Feature & $p$ value\\
		\hline
		coeff. of var. of 11th LPCC & $1.0832\cdot10^{-04}$\\
		position of max. of GNE & $1.3509\cdot10^{-04}$\\
		index of dispersion of 11th LPCC & $1.6228\cdot10^{-04}$\\
		min. of MPSD & $3.0797\cdot10^{-04}$\\
		1st percentile of MPSD & $3.1013\cdot10^{-04}$\\
		modulation of MPSD & $3.1013\cdot10^{-04}$\\
		relative range of MPSD & $3.1013\cdot10^{-04}$\\
		harmonic mean of MPSD & $3.1085\cdot10^{-04}$\\
		modulation of $\mbox{GQ}_{\mathrm{open}}$ & $4.0151\cdot10^{-04}$\\
		relative range of $\mbox{GQ}_{\mathrm{open}}$ & $4.0151\cdot10^{-04}$\\
		\hline
		\hline
		\end{tabular}
\end{table}

\begin{table}[ht!]
		\caption{10 most significant features selected by Mann-Whitney U~test in scenario C2: PARCZ, males (UCPP\,--\,Unsmooth Cepstral Peak Prominence, LPCT\,--\,Linear Predictive Cosine Transform coefficients, CMS\,--\,Cepstral Mean Subtraction coefficients, $\mbox{IMF-NSR}_\mathrm{RE}$\,--\,Noise-to-Signal Ratio derived from IMF based on second-order R{\'e}nyi Entropy, AE(Laplacian)\,--\,Approximate Entropy based on Laplacian kernel).}
		\label{tab:sign_C2}
		\footnotesize
		\centering
		\begin{tabular}{l l}
		\hline
		\hline
		Feature & $p$ value\\
		\hline
		slope of UCPP & $3.4500\cdot10^{-05}$\\
		40th percentile of 2nd LPCT & $4.2438\cdot10^{-05}$\\
		offset of linear regression of 5th CMS & $4.2438\cdot10^{-05}$\\
		1st quartile of 2nd LPCT & $5.2089\cdot10^{-05}$\\
		offset of linear regression of $\mbox{IMF-NSR}_\mathrm{RE}$ & $6.3797\cdot10^{-05}$\\
		mean excluding 40\,\% outliers of 2nd LPCT & $7.0546\cdot10^{-05}$\\
		mean excluding 50\,\% outliers of 2nd LPCT & $7.0546\cdot10^{-05}$\\
		offset of linear regression of AE (Laplacian) & $7.0546\cdot10^{-05}$\\
		mean excluding 10\,\% outliers of 2nd LPCT & $7.7966\cdot10^{-05}$\\
		mean excluding 20\,\% outliers of 2nd LPCT & $7.7966\cdot10^{-05}$\\
		\hline
		\hline
		\end{tabular}
\end{table}

\begin{table}[ht!]
		\caption{10 most significant features selected by Mann-Whitney U~test in scenario C3: PARCZ, all (LFCC\,--\,Linear Frequency Cepstral Coefficients, MPSD\,--\,Median of Power Spectral Density, FADFA\,--\,Fluctuation Amplitudes of Detrended Fluctuation Analysis, ICC\,--\,Inferior Colliculus Coefficients).}
		\label{tab:sign_C3}
		\footnotesize
		\centering
		\begin{tabular}{l l}
		\hline
		\hline
		Feature & $p$ value\\
		\hline
		interquartile range of 18th LFCC & $1.8665\cdot10^{-05}$\\
		median absolute deviation of 18th LFCC & $1.9509\cdot10^{-05}$\\
		min. of MPSD & $1.9970\cdot10^{-05}$\\
		modulation of MPSD & $2.0055\cdot10^{-05}$\\
		relative range of MPSD & $2.0055\cdot10^{-05}$\\
		harmonic mean of MPSD & $2.1543\cdot10^{-05}$\\
		40th percentile of FADFA & $3.7451\cdot10^{-05}$\\
		median absolute deviation of 2nd ICC & $5.7179\cdot10^{-05}$\\
		mean excluding 10\,\% outliers of 2nd ICC & $5.9621\cdot10^{-05}$\\
		mean excluding 20\,\% outliers of 2nd ICC & $5.9621\cdot10^{-05}$\\
		\hline
		\hline
		\end{tabular}
\end{table}

To summarize the discussion about the databases we can say that the MEEI database should no longer be used as a~benchmark. The more challenging one is PARCZ database, however it has some disadvantages from the scientific point of view: it is in Czech language, which is not a~widespread language in the world; it is focused only on PD people with hypokinetic dysarthria; and it is not really suitable for a~binary classification (at least 4 classes should be considered). Probably the most suitable database for the evaluation of pathological speech identification methods is PdA. The classification accuracies are still challenging and they can be significantly improved. Another advantage of this database is that it is freely available for research purposes. The only disadvantage is the limitation to Spanish language. This should not be such a~big problem in the case of sustained vowel [a], but there will certainly be cultural differences when dealing with the spoken text analysis.

Another question is, whether analysis of vowel [a] is really the best way to identify pathological speech, at least in the field of vowels analysis (not considering the other speech tasks like read/repeated/spontaneous words, sentences, etc.). Although most of researchers automatically use sustained vowel [a], just a~few publications report classification accuracies based on analysis of the other vowels. For this purpose Henriquez et al. made an experiment where they tried to identify pathological speech based on analysis of 5 Spanish vowels separately ([a], [e], [i], [o], [u])~\cite{Henriquez2009}. They observed that in comparison to the other vowels the classification based on vowel [a] provides slightly better results. Probably the choice of vowel [a] is good when classifying the pathological speech generally (but still this should be proved by robust testing in future). However, as soon as we focus on a~specific pathology, we can get better results when analysing another vowel. For instance Orozco-Arroyave et al. classified hypokinetic dysarthria in patients with Parkinson's disease and found out that it is more advantageous to analyse vowel [o]~\cite{Orozco2013}.

Our initial idea was also to try the inter-database classification (training the classifier on one database and testing it using the other one). However the classification results were very poor. It was caused mainly by these two facts: 1) When we trained the classifier on MEEI database, the features reflecting the signal length were selected. But these features were not so significant when testing them on PdA or PARCZ databases. 2) The PARCZ database is very different from MEEI or PdA, because it contains only one specific speech disorder (hypokinetic dysarthria), while the other two databases contain many different voice pathologies.

\begin{table}[ht!]
		\caption{Best significance levels (computed using the Mann-Whitney U~test) selected for all 36 features originally introduced in this work (F\,--\,female, M\,--\,male, MF\,--\,all genders).}
		\label{tab:results_introduced}
		\scriptsize
		\centering
		\begin{tabular}{p{.38\textwidth} l l c c c}
		\hline
		\hline
		Local feature & High-level feature & $p$ value & Sc. ID & Dataset & Gender\\
		\hline
		IMF-CPP (Cepstral Peak Prominence extracted from the 1st IMF) & error of linear regression & $6.9443\cdot10^{-32}$ & P3 & PdA & MF\\
		MFP (Modulation Frequency of Peak) & - & $3.2291\cdot10^{-28}$ & M3 & MEEI & MF\\
		RPHM (Relative Peak Height of Modulation spectra) & - & $1.5614\cdot10^{-27}$ & M3 & MEEI & MF\\
		MSER (Modulation Spectra Energy Ratio) & - & $2.9477\cdot10^{-26}$ & M3 & MEEI & MF\\
		BCPD (BiCepstral Phase Distance) & harmonic mean & $4.9663\cdot10^{-19}$ & M3 & MEEI & MF\\
		HFEBC (High Frequency Energy of one-dimensional BiCepstral index) & - & $2.5381\cdot10^{-15}$ & M3 & MEEI & MF\\
		LFEBC (Low Frequency Energy of one-dimensional BiCepstral index) & - & $2.5381\cdot10^{-15}$ & M3 & MEEI & MF\\
		AE (triangular kernel) & offset of linear regression & $7.7587\cdot10^{-15}$ & P3 & PdA & MF\\
		BCII (BiCepstral Index Interference) & - & $9.1669\cdot10^{-15}$ & P3 & PdA & MF\\
		AE (exponential kernel) & 1st quartile & $1.4641\cdot10^{-14}$ & P3 & PdA & MF\\
		SE (exponential kernel) & median & $5.5444\cdot10^{-14}$ & P3 & PdA & MF\\
		IMF-GNE (Glottal-to-Noise Excitation ratio based on the 1st IMF) & mean & $6.0985\cdot10^{-14}$ & M3 & MEEI & MF\\
		AE (Cauchy kernel) & mean & $1.3344\cdot10^{-13}$ & M3 & MEEI & MF\\
		AE (spherical kernel) & offset of linear regression & $1.4599\cdot10^{-13}$ & P3 & PdA & MF\\
		SE (spherical kernel) & mean excluding 40\,\% outliers & $2.5551\cdot10^{-13}$ & P3 & PdA & MF\\
		$\mbox{IMF-SNR}_\mathrm{RE}$ (based on second-order R{\'e}nyi Entropy) & mean & $3.1582\cdot10^{-12}$ & P3 & MEEI & MF\\
		LCBCER (Low Cepstra/BiCepstra Energy Ratio) & mean & $5.6242\cdot10^{-12}$ & M3 & MEEI & MF\\
		IMF-FD (based on Fractal Dimension) & median & $1.4092\cdot10^{-11}$ & M3 & MEEI & MF\\
		ICER (Inferior Colliculus Energy Ratio) & - & $2.8611\cdot10^{-11}$ & M3 & MEEI & MF\\
		BMD (Bispectral Module Distance) & median & $5.3843\cdot10^{-11}$ & M3 & MEEI & MF\\
		AE (circular kernel) & 4th moment & $7.3659\cdot10^{-10}$ & P3 & PdA & MF\\
		SE (circular kernel) & std & $1.3186\cdot10^{-09}$ & P3 & PdA & MF\\
		SE (triangular kernel) & mean & $1.3839\cdot10^{-09}$ & P2 & PdA & M\\
		BPD (Bispectral Phase Distance) & mean & $1.0753\cdot10^{-08}$ & P3 & PdA & MF\\
		$\mbox{IMF-NSR}_\mathrm{RE}$ (based on second-order R{\'e}nyi Entropy) & 90th percentile & $1.1298\cdot10^{-08}$ & P2 & PdA & M\\
		$\mbox{IMF-SNR}_\mathrm{ZCR}$ (based on Zero-Crossing Rate) & mean & $1.5834\cdot10^{-08}$ & M3 & MEEI & MF\\
		HCBCER (High Cepstra/BiCepstra Energy Ratio) & 1st percentile & $4.7190\cdot10^{-08}$ & P3 & PdA & MF\\
		SE (Cauchy kernel) & harmonic mean & $5.1617\cdot10^{-08}$ & M1 & MEEI & F\\
		BCPII (BiCepstrum Phase Interference Index) & - & $1.7589\cdot10^{-07}$ & M1 & MEEI & F\\
		RPHIC (Relative Peak Height of Inferior Colliculus) & - & $2.2653\cdot10^{-07}$ & M2 & MEEI & M\\
		BCMD (BiCepstral Module Distance) & harmonic mean & $1.7894\cdot10^{-06}$ & M3 & MEEI & MF\\
		HSBER (High Spectra/Bispectra Energy Ratio) & mean & $3.2552\cdot10^{-06}$ & M1 & MEEI & F\\
		AE (Laplacian kernel) & offset of linear regression & $4.2554\cdot10^{-06}$ & P2 & PdA & M\\
		SE (Laplacian kernel) & 80th percentile & $7.5456\cdot10^{-06}$ & P2 & PdA & M\\
		BCMII (BiCepstrum Module Interference Index) & - & $1.1755\cdot10^{-04}$ & P3 & PdA & MF\\
		LSBER (Low Spectra/Bispectra Energy Ratio) & mean excluding 20\,\% outliers & $3.7662\cdot10^{-04}$ & M2 & MEEI & M\\
		\hline
		\hline
		\end{tabular}
\end{table}

From the feature selection point of view it is interesting to point out that many segmental parameters have been found significant. Some of them (MSC, ACW, LPCC, LPCT, CMS and ICC) were also selected as the 10 most significant ones. Although the segmental features are not very frequent when analysing pathological speech (except MFCC, MSC and ICC), their potential seems to be high. In fact, to our best knowledge, features like MFCCE, LFCC, CMS or ACW were used for this purpose for the first time in the present research.

In this work we have introduced 36 new speech features. Just one (IMF-CPP) has been mentioned among the 10 most significant ones (Table\,\ref{tab:sign_P2} and \ref{tab:sign_P3}), however the rest of them are significant ($p < 0.05$) as well, see Table\,\ref{tab:results_introduced}. In the case of MEEI database we firstly checked if high-level features do not reflect the length of signal. If not we kept the feature in the table. If yes we checked the next most significant variant of the local parameter. At the top positions of this table we can see mainly features based on modulation spectra, bicepstrum and approximate entropy.

\section{Conclusions}
\label{sec:conclusions}

This work provides an insight into the robust and complex approach of pathological speech analysis. To our best knowledge this is the first contribution providing a complex evaluation of feature significance from different fields of speech signal processing (e.\,g. speech analysis, recognition, coding, enhancement etc.). It is also the first contribution deriving conclusions according to robust tests where 3 (English, Spanish, Czech) databases were used. These languages belong to 3 different language groups (Germanic, Romanic, Slavic). In general, the work has 4 goals, yet each of them has its conclusion.

1) \emph{According to complex parameterization and consequent robust testing identify features that have the largest discriminative power in the field of pathological speech analysis.} Unfortunately most of the works published in the field of pathological speech analysis provides conclusions based on a~limited set of parameterization methods. In other words there is still a~lack of publications providing a~complex overview of features quantifying pathological speech and providing strong conclusions supported by a~robust testing.

Our work is unique in this way, because together it provides testing based on 128 local features. We made the wide overview of all these features used for pathological speech quantification so that the researchers can find out what exactly the specific feature quantifies and how it can be implemented. We used features describing phonation, tongue movement, speech quality (including non-linear dynamic features and features based on bispectrum/bicepstrum, empirical mode decomposition, wavelet decomposition) and segmental features.

Using the non-parametric Mann-Whitney~U test we observed that among all parameterization techniques those based on segmental features provide the best classification results. To our best knowledge this is the first work that tested the significance of 11 segmental features. The largest discriminative power can be obtained thanks to especially segmental features like modulation spectra coefficients, adaptive component weighted coefficients and linear predictive cepstral coefficients. Although the segmental features are not very frequent when analysing pathological speech, their potential seems to be high. However, their disadvantage is that they are usually difficult to be interpreted clinically. This is probably the reason they are not frequently used. Although they can provide good classification results they don't say much about specific pathology or speech dysfunction.

2) \emph{Design new features that can quantify hoarseness, breathiness and non-linearities in pathological speech signals}. Clinical signs of vocal fold dysfunctions are usually associated with breathiness or hoarseness. Moreover voice can become aperiodic, noisy-like, and it is very difficult to find any regularities in the signal. Sometimes there is a~frequent presence of sub-harmonics and chaos, which can lead to a~failure of conventional techniques of speech signal analysis and which requires new parameterization methods developed specifically for pathological speech description.

We introduced 36 new measures based on modulation spectra, inferior colliculus coefficients, bicepstrum, sample and approximate entropy and empirical mode decomposition. Features based on modulation spectra quantify instability of vocal fold vibrations and complements features based on inferior colliculus coefficients that reflect the misplacement of articulators. Due to incorrect glottal closure the pathological voice contains much more white noise that can be effectively quantified by proposed features derived from bicepstrum. New features based on empirical mode decomposition are able to describe the noise component of analysed signal as well. Finally we proposed different kernel-based approximate and sample entropies to measure regularities inside the signal.

These novel features were statistically processed by the non-parametric Mann-Whitney~U test. All of them have been identified as significant and they have passed the feature selection process in at least one database. Moreover, some of them were listed among top ten significant features selected in specific scenario (they outperformed the other conventional features). In other words they helped to improve classification results in terms of accuracy, sensitivity and specificity and due to their effective quantification abilities they have high impact on the future work.

3) \emph{Prove that the proposed large set parameterization approach can provide better classification results (with respect to classification accuracy, sensitivity and specificity) than those published in the field of pathological speech analysis by the other researchers}. We tested the significance of all the mentioned parameters on 3 (English, Spanish, Czech) databases. In the case of the Massachusetts Eye and Ear Infirmary (MEEI) database we get accuracy, sensitivity and specificity equal to $100.0\pm0.0\,\%$, which are the best results that have been published in the frame of this database (Henriquez et al. reached $99.69\,\%$~\cite{Henriquez2009}). However, we are very critical with these results. Therefore we discussed their trustability and the viability of using the MEEI database as a~benchmark for pathological speech signal analysis. 

The results obtained with the PdA database are more challenging. When we considered the single-classifier approach, we reached the accuracy $82.1\pm3.3\,\%$, which is the best among the published numbers (Arias-Londono et al. published $81.7\,\%$~\cite{Londono2011}).

Regarding the last Czech Parkinsonian Speech Database (PARCZ), we obtained poor accuracy ($67.9\pm6.0\,\%$), however, this is probably due to the fact that we used a~binary classifier for a~multi-class database. We would like to do more experiments with this database in a~near future splitting the data into healthy speech and mild, moderate and severe dysarthria.

4) \emph{Select a~database that has high potential for the future, especially in terms of speech features design, tuning and testing}. Due to some issues related to vowels' length, different sampling frequencies, different recording conditions, etc., the MEEI database should no longer be used as a~benchmark. Results obtained using this database are not trustable. The more challenging one is PARCZ database. Unfortunately it is focused only on parkinsonic people with hypokinetic dysarthria and it is not really suitable for a~binary classification (at least 4 classes should be considered).

Therefore the most suitable database for the evaluation of pathological speech identification methods is PdA, where the classification accuracies are still challenging and they can be significantly improved. Moreover this database is freely available for research purposes.\\

There are many works that deal with the development of pathological voice identification methods and there is still a~lot that can be improved in this field of science. However, the researchers should go further and focus not only on the pathological speech identification, but also on more sophisticated analysis that would be more helpful for doctors and that can make the treatment or diagnosis more effective. Probably the most important challenges to face in the next decade are: 
\begin{enumerate}
	\item \emph{Identification of particular voice pathology.} Identification of pathological speech itself is not so interesting. There are many pathologies (adductor spasmodic dysphonia, erythema, hypokinetic dysarthria, etc.) and the issue is to classify them individually. This can be very problematic, therefore we propose to do some kind of clustering and split this big set into subsets, for example according to the way they are reflected in speech (problems with tongue movement, improper work of soft palate, disordered vocal folds, etc.).
	\item \emph{Identification of voice pathology in its first stage or estimation of its progress.} This would enable doctors to start the treatment very early and slow down the progress.
\end{enumerate}

We are going to deal with these issues in future works. However this research is very dependent on good databases and especially in the case of voice pathology identification, as in its first stage, there is still a~lack of suitable training data.

\section*{Acknowledgments}

This work was supported by project NT13499 (Speech, its impairment and cognitive performance in Parkinson's disease), GACR 102/12/1104, COST IC1206 and project ``CEITEC, Central European Institute of Technology'': (CZ.1.05/1.1.00/02.0068) from the European Regional Development Fund, FEDER and Ministerio de Econom\'{i}a y Competitividad TEC2012-38630-C04-03 and -04 (Kingdom of Spain). The described research was performed in laboratories supported by the SIX project; the registration number CZ.1.05/2.1.00/03.0072, the operational program Research and Development for Innovation.

\section*{References}

\bibliography{Mekyska_neurocomputing}

\end{document}